\documentclass{aa}
\usepackage{epsf,epsfig}
\newcommand{\Chi}{\raisebox{0.4ex}{$\chi\,$}}

\begin{document}
 
\title{Distribution of star-forming complexes in dwarf irregular galaxies}
\titlerunning{Distribution of bright lumps in irregular dwarfs}

\author{Bernhard R. Parodi \and Bruno Binggeli}
\authorrunning{Parodi \& Binggeli}  
 
\offprints{B. R. Parodi}    
 
\institute{Astronomisches Institut der Universit\"at Basel, 
           Venusstrasse 7, CH-4102 Binningen, Switzerland\\
           \email{parodi@astro.unibas.ch, binggeli@astro.unibas.ch}          
}

\date{Received / Accepted} 
 
\abstract{We study the distribution of bright star-forming complexes
in a homogeneous sample of 72 late-type (``irregular'') dwarf galaxies
located within the 10 Mpc volume. Star-forming complexes are
identified as bright lumps in $B$-band galaxy images and isolated by
means of the unsharp-masking method. For the sample as a whole
the radial number distribution of bright lumps  largely traces the
underlying exponential-disk light profiles, but peaks at a 10
percent smaller scale length. Moreover, the presence of a tail of star
forming regions out to at least six optical scale lengths provides
evidence against a systematic star formation truncation within
that galaxy extension. Considering these findings, we apply a scale
length-independent concentration index, taking into account the
implied non-uniform random spread of star formation regions throughout
the disk. The number profiles frequently manifest a second, minor peak
at about two scale lengths. Relying on a two-dimensional stochastic
self-propagating star formation model, we show these secondary peaks
to be consistent with triggered star formation; for a few of the
brighter galaxies a peculiar peak distribution is observed that is
conceivably due to the onset of shear provided by differential
rotation.  On scales between 100 and 1000 pc, and by taking into
account exponential-disk structure, bright lumps reveal cluster
dimensions between 1.3 and 2, with a weak trend to higher dimensions
for brighter galaxies. Cluster dimension weakly anticorrelates with
the lumpiness index (the fraction of the total galaxy light due to the
light contributed by the lumps), the latter index showing no
dependence on luminosity. Lump spreading within the disk, as measured
by the concentration index,  and lump clustering, as given by the
cluster dimension, are not linked to each other. Interpreting cluster
dimension in terms of porosity of a self-similar intragalactic medium,
we derive a relation between current star formation rate, scale
length, and porosity.  \keywords{ galaxies: irregular -- galaxies:
morphology --  galaxies: star formation} }

\maketitle

\section{Introduction}

\begin{table*}[btp]
\caption[]{Summary of terms}
\label{summary}
     $$ 
     \begin{array}{p{0.2\linewidth}p{0.76\linewidth}}
\hline\hline
\noalign{\smallskip}
Lumps: & bright residual features seen in galaxy images after a median filtered version is subtracted from\\
       &  the original image; synonymous expressions: bright spots or knots;  physical correspondents: star-\\
       & forming complexes encompassing H$\;$II regions, young star clusters, and stellar associations;\\
Lumpiness index $\Chi$:   & fractional flux or ratio of the flux due to the lumps within the residual image and the total galaxy\\
                          & light of the original image.\\ 
Cluster dimension $D$:    & correlation dimension for a discrete set of lump centers in a plane, i.e. within a radius $r$ around\\
                          & a typical lump there are $n \propto r^D$ other lumps; no weighting for lump size or luminosity. If consi-\\
                          & dered as an indirect measure for a three dimensional medium's fractal dimension, $D$ may be\\
                          & related to the volume filling factor of the empty regions, called porosity.\\
Concentration index $CI$: & concentration index as the ratio of lump centers in an inner circle and lump centers in an outer\\
                          & annulus, normalized according to some prescription; no weighting for lump size or luminosity.\\
\noalign{\smallskip}
\hline\hline
     \end{array}
     $$ 
\end{table*}
\begin{table*}[tbp] 
  \caption[t]{Galaxy and bright-lump data for the 72 irregular dwarf galaxies of our sample} 
  \scriptsize 
  $$  
  \begin{array}{llrcrrrccrcccccc} 
    \hline 
    \noalign{\smallskip} 
    {\rm galaxy}&{\rm type} & M_B &\mu_0^B & R_d & R_{25} &{\rm b/a} & v_{\rm rot} & {\rm Ref.}^\dagger & 
    N & R_l/R_d & R_{1st}/R_d & R_{2nd}/R_d & CI(R_{25}) & D & \Chi \\ 
    \noalign{\smallskip} 
    \;\;\;\;(1) & (2) & (3)\;\; & (4) & (5) & (6) & (7) & (8) & (9) & (10) & (11) & (12) & (13) & (14) & (15) & (16)\\ 
    \noalign{\smallskip} 
    \hline 
    \noalign{\smallskip} 
    {\rm DDO53}         & {\rm Im}      & -12.84 & 23.06 & 281 & 544 & 0.87 & \dots & I   &  12 & 1.12 & 1.5 & 3.0 & 0.37 &\dots& 0.147 \\ 
    {\rm UGC4483}       & {\rm Im}      & -12.66 & 22.15 & 182 & 578 & 0.50 & 22    & I   &   8 &\dots & 1.0 & 0.0 & 1.26 &\dots& 0.101 \\ 
    {\rm Kar54}         & {\rm Im}      & -14.68 & 23.89 &1283 &1414 & 0.85 & 28    & I   &   1 &\dots &\dots&\dots&\dots &\dots& 0.019 \\ 
    {\rm UGC4998}       & {\rm Im}      & -15.78 & 22.50 & 843 &1879 & 0.50 & \dots & I   &  10 & 1.14 &\dots&\dots& 1.13 &\dots& 0.058 \\ 
    {\rm HoI}           & {\rm Im}      & -15.44 & 22.75 &1007 &2089 & 0.83 & 28    & I   &  85 & 0.90 & 2.0 & 1.0 & 0.81 & 1.30& 0.180 \\ 
    {\rm BK1N}          & {\rm Im}      & -12.70 & 24.49 & 529 &1832 & 0.42 & \dots & I   &   0 &\dots &\dots&\dots&\dots &\dots& 0.035 \\ 
    {\rm NGC2976}       & {\rm Sd}      & -17.37 & 20.52 & 807 &2988 & 0.42 & 54    & I   & 218 & 0.36 & 2.0 &\dots& 0.49 & 1.59& 0.077 \\ 
    {\rm UGC5423}       & {\rm BCD}     & -13.76 & 22.30 & 315 & 694 & 0.67 & 27    & I   &   4 &\dots &\dots&\dots&\dots &\dots& 0.051 \\ 
    {\rm DDO82}         & {\rm Sm/BCD}  & -14.75 & 22.73 & 593 &1209 & 0.56 & \dots & I   &   5 &\dots &\dots&\dots&\dots &\dots& 0.045 \\ 
    {\rm DDO87}         & {\rm Im}      & -14.18 & 24.13 & 929 & 746 & 1.00 & 66    & I   &   7 &\dots & 1.5 & 4.5 &\dots &\dots& 0.069 \\ 
    {\rm Kar73}         & {\rm Im}      & -10.81 & 24.53 & 228 & 120 & 0.67 & \dots & I   &   1 &\dots &\dots&\dots&\dots &\dots& 0.030 \\ 
    {\rm UGC8215}       & {\rm Im}      & -12.71 & 22.29 & 234 & 593 & 0.74 & 14    & III &   0 &\dots &\dots&\dots&\dots &\dots& \dots \\ 
    {\rm DDO167}        & {\rm Im}      & -12.34 & 22.64 & 248 & 565 & 0.60 & 17    & III &  10 & 0.78 & 1.5 &\dots& 0.55 &\dots& 0.083 \\ 
    {\rm DDO168}        & {\rm Im}      & -14.97 & 21.89 & 715 &2048 & 0.40 & 31    & III &  70 & 0.58 & 1.0 & 2.0 & 1.47 & 1.31& 0.110 \\ 
    {\rm DDO169}        & {\rm Im}      & -15.30 & 22.15 &1032 &2692 & 0.32 & 26    & III &  15 & 0.74 & 1.0 &\dots& 1.92 &\dots& 0.097 \\ 
    {\rm UGC8508}       & {\rm Im}      & -13.69 & 21.24 & 293 &1005 & 0.56 & 26    & III &  33 & 0.96 & 2.0 &\dots& 0.63 & 1.82& 0.042 \\ 
    {\rm NGC5229}       & {\rm Sd}      & -14.44 & 20.51 & 482 &2007 & 0.19 & 56    & III &  30 & 1.10 & 1.0 &\dots& 1.56 & 1.56& 0.120 \\ 
    {\rm NGC5238}       & {\rm Sdm}     & -15.03 & 21.90 & 544 &1507 & 0.68 & 19    & III &  22 & 0.46 & 0.5 &\dots& 6.27 & 1.56& 0.086 \\ 
    {\rm DDO181}        & {\rm Im}      & -13.30 & 22.00 & 372 &1064 & 0.44 & 21    & III &  44 & 1.11 & 0.5 & 3.5 & 0.99 & 1.67& 0.207 \\ 
    {\rm DDO183}        & {\rm Im}      & -13.90 & 21.89 & 337 &1016 & 0.75 & 16    & III &   7 &\dots &\dots&\dots&\dots &\dots& 0.035 \\ 
    {\rm UGC8833}       & {\rm Im}      & -11.95 & 22.41 & 177 & 425 & 0.71 & 19    & III &   2 &\dots &\dots&\dots&\dots &\dots& 0.020 \\ 
    {\rm HoIV}          & {\rm Im}      & -15.95 & 22.11 &1634 &4442 & 0.26 & 39    & III &  65 & 1.06 & 1.0 & 4.0 & 0.71 & 1.78& 0.175 \\ 
    {\rm NGC5474}       & {\rm Scd}     & -17.52 & 20.75 & 992 &3741 & 0.96 & 41    & III & 198 & 1.41 & 1.5 & 4.0 & 0.50 & 1.55& 0.109 \\ 
    {\rm NGC5477}       & {\rm Sm}      & -15.24 & 21.59 & 559 &1763 & 0.70 & 32    & III &  37 & 1.01 & 2.0 & 1.0 & 0.49 & 1.61& 0.121 \\ 
    {\rm DDO190}        & {\rm Im}      & -15.17 & 20.99 & 382 &1393 & 0.88 & 44    & III &  24 & 1.02 & 1.0 & 0.0 & 1.23 & 1.49& 0.030 \\ 
    {\rm DDO194}        & {\rm Im}      & -15.00 & 23.03 & 981 &1814 & 0.64 & 38    & III &   0 &\dots &\dots&\dots&\dots &\dots& \dots \\ 
    {\rm UGC6541}       & {\rm Sm/BCD}  & -13.40 & 22.49 & 294 & 677 & 0.53 & 17    & IV  &   7 &\dots & 0.5 &\dots&\infty&\dots& 0.037 \\ 
    {\rm NGC3738}       & {\rm Irr}     & -15.80 & 22.40 & 685 &1544 & 0.79 & 69    & IV  &  74 & 0.28 & 0.0 &\dots& 12.3 & 1.95& 0.062 \\ 
    {\rm NGC3741}       & {\rm Im/BCD}  & -13.34 & 21.71 & 207 & 609 & 0.77 & 38    & IV  &  14 & 1.04 & 1.0 & 2.0 & 1.95 &\dots& 0.029 \\ 
    {\rm DDO99}         & {\rm Im}      & -14.51 & 23.09 & 904 &1560 & 0.47 & 23    & IV  &  70 & 0.83 & 0.5 & 2.5 & 1.85 & 1.69& 0.132 \\ 
    {\rm NGC4068}       & {\rm Sm/BCD}  & -15.69 & 22.89 & 950 &2089 & 0.57 & 30    & IV  & 111 & 0.50 & 1.5 &\dots& 0.64 & 1.34& 0.243 \\ 
    {\rm NGC4163}       & {\rm BCD}     & -14.06 & 22.68 & 422 & 597 & 0.66 & 17    & IV  &  17 & 1.14 & 0.5 &\dots& 1.88 &\dots& 0.076 \\ 
    {\rm UGC7298}       & {\rm Im}      & -13.71 & 22.44 & 413 & 959 & 0.67 & 19    & IV  &   0 &\dots &\dots&\dots&\dots &\dots& \dots \\ 
    {\rm NGC4248}       & {\rm IBm}     & -16.33 & 21.57 &1066 &3313 & 0.43 & 42    & IV  &  16 & 1.02 & 0.5 &\dots&\infty&\dots& 0.072 \\ 
    {\rm DDO127}        & {\rm Sm}      & -14.39 & 24.72 &1003 &1455 & 0.51 & 32    & IV  &   3 &\dots &\dots&\dots&\dots &\dots& 0.039 \\
    {\rm UGC7639}       & {\rm dS0/BCD} & -15.58 & 22.57 &1015 &2184 & 0.54 & 25    & IV  &  64 & 0.54 & 0.5 &\dots& 4.24 & 1.48& 0.041 \\ 
    {\rm UGC288}        & {\rm Im}      & -13.82 & 22.98 & 441 & 799 & 0.45 & 25    & VI  &   1 &\dots &\dots&\dots&\dots &\dots& 0.000 \\ 
    {\rm UGC685}        & {\rm Im/BCD}  & -14.92 & 21.96 & 509 &1392 & 0.63 & 38    & VI  &   1 &\dots &\dots&\dots&\dots &\dots& 0.019 \\ 
    {\rm UGC1281}       & {\rm Sd}      & -15.83 & 20.90 &1125 &3725 & 0.16 & 50    & VI  &  94 & 0.70 & 0.5 & 2.0 & 1.96 & 1.69& 0.140 \\ 
    {\rm NGC1156}       & {\rm IBm}     & -17.68 & 20.31 & 922 &3971 & 0.62 & 55    & VI  &  63 & 0.99 & 1.0 & 2.0 & 0.98 & 1.80& 0.118 \\ 
    {\rm UGC2684}       & {\rm Im}      & -13.13 & 23.30 & 618 &1100 & 0.38 & 37    & VI  &   1 &\dots &\dots&\dots&\dots &\dots& 0.055 \\ 
    {\rm UGC2716}       & {\rm Sm}      & -15.08 & 23.24 &1047 &1673 & 0.47 & 27    & VI  &   7 &\dots & 0.0 &\dots&\infty&\dots& 0.044 \\ 
    {\rm UGC2905}       & {\rm Im}      & -14.41 & 21.30 & 277 & 889 & 0.63 & 26    & VI  &   1 &\dots &\dots&\dots&\dots &\dots& 0.016 \\ 
    {\rm UGC3303}       & {\rm Sd}      & -15.90 & 23.07 &1718 &3031 & 0.61 & 79    & VI  &   7 &\dots &\dots&\dots&\dots &\dots& 0.048 \\ 
    {\rm PGC17716}      & {\rm SBd}     & -16.79 & 21.90 &1310 &3207 & 0.76 & 51    & VI  &   5 &\dots &\dots&\dots& 4.11 &\dots& 0.068 \\ 
    {\rm A0554+07}      & {\rm Im}      & -12.25 & 23.46 & 254 & 366 & 0.97 & 27    & VI  &   0 &\dots &\dots&\dots&\dots &\dots& \dots \\ 
    {\rm UGC3476}       & {\rm Im}      & -14.26 & 21.54 & 484 &1477 & 0.26 & 47    & VI  &   5 &\dots &\dots&\dots& 3.18 &\dots& 0.082 \\ 
    {\rm UGC3600}       & {\rm Im}      & -13.53 & 22.88 & 672 &1315 & 0.29 & 39    & VI  &   2 &\dots &\dots&\dots&\dots &\dots& 0.049 \\ 
    {\rm NGC2337}       & {\rm IBm/BCD} & -16.39 & 21.59 & 801 &2637 & 0.62 & 74    & VI  &  33 & 0.91 & 1.0 & 2.0 & 2.15 & 1.66& 0.094 \\ 
    {\rm UGC4115}       & {\rm Im}      & -13.51 & 22.66 & 493 &1054 & 0.44 & 44    & VI  &  16 & 0.84 & 1.5 & 3.5 & 0.00 &\dots& 0.077 \\ 
    {\rm NGC2537}       & {\rm Sm/BCD}  & -16.60 & 22.10 & 711 &1911 & 0.91 & 102   & VI  &  53 & 0.47 & 1.0 &\dots& 3.34 & 1.59& 0.079 \\ 
    {\rm DDO64}         & {\rm Im}      & -13.95 & 22.56 & 636 &1532 & 0.39 & 42    & VI  &  17 & 1.83 & 1.5 &\dots& 0.65 &\dots& 0.130 \\ 
    {\rm ESO 473-G024}  & {\rm Im}      & -13.74 & 22.00 & 454 &1179 & 0.46 & 17    & VII &  22 & 0.34 & 1.5 &\dots& 0.10 & 1.45& 0.196 \\ 
    {\rm ESO 115-G021}  & {\rm Sm}      & -15.18 & 21.63 &1164 &3455 & 0.15 & 50    & VII & 372 & 0.85 & 1.0 &\dots& 1.24 & 1.55& 0.257 \\ 
    {\rm ESO 154-G023}  & {\rm Sm}      & -16.23 & 20.10 &1141 &5635 & 0.22 & 56    & VII & 318 & 1.54 & 2.0 & 3.5 & 0.39 & 1.62& 0.204 \\ 
    {\rm NGC 1311}      & {\rm Sm}      & -15.69 & 21.74 & 884 &2587 & 0.30 & 42    & VII &  29 & 0.56 & 0.5 &\dots& 4.02 & 1.78& 0.107 \\ 
    {\rm IC 1959}       & {\rm Sdm}     & -15.92 & 19.25 & 563 &2485 & 0.24 & 58    & VII & 225 & 2.29 & 2.5 & 0.5 & 0.28 & 1.56& 0.142 \\ 
    {\rm IC 2038}       & {\rm Sd}      & -14.47 & 21.63 & 596 &1931 & 0.29 & 33    & VII &  33 & 0.41 & 0.5 &\dots& 4.10 & 1.76& 0.066 \\ 
    {\rm NGC 1800}      & {\rm Sm/BCD}  & -16.25 & 21.64 & 683 &2086 & 0.57 & 39    & VII &  59 & 0.55 & 0.5 &\dots& 7.78 & 1.60& 0.083 \\ 
    {\rm AM 0521-343}   & {\rm Im}      & -14.24 & 21.60 & 358 &1020 & 0.67 & 42    & VII &  11 & 1.14 & 2.0 & 1.0 & 0.49 &\dots& 0.058 \\ 
    {\rm ESO 555-G028}  & {\rm Im}      & -13.79 & 23.43 & 870 &1399 & 0.37 & 54    & VII &   7 &\dots & 1.0 &\dots& 1.11 &\dots& 0.065 \\ 
    {\rm ESO 489-G056}  & {\rm Im}      & -12.30 & 23.00 & 186 & 377 & 0.72 & 19    & VII &  22 & 0.78 & 1.0 &\dots& 0.82 & 1.47& 0.094 \\ 
    {\rm ESO 490-G017}  & {\rm Im}      & -14.18 & 21.30 & 269 & 918 & 0.79 & 28    & VII &  36 & 0.90 & 0.5 & 1.5 & 4.21 & 1.47& 0.076 \\ 
    {\rm ESO 308-G022}  & {\rm Im}      & -13.71 & 23.81 & 609 & 683 & 0.90 & 33    & VII &  11 & 1.26 & 0.5 & 2.0 & 0.46 &\dots& 0.117 \\ 
    {\rm PGC 20125}     & {\rm Im}      & -13.46 & 23.75 & 676 & 502 & 0.70 & 36    & VII &  61 & 0.96 & 1.5 & 0.0 & 2.16 & 1.29& 0.292 \\ 
    {\rm ESO 558-PN011} & {\rm Im}      & -16.30 & 22.08 & 825 &2339 & 0.71 & 68    & VII &  49 & 0.25 & 1.0 &\dots& 18.7 & 1.38& 0.120 \\ 
    {\rm ESO 059-G001}  & {\rm Im}      & -14.49 & 22.20 & 536 &1288 & 0.71 & 46    & VII &  60 & 0.70 & 1.0 &\dots& 1.20 & 1.70& 0.149 \\ 
    {\rm ESO 006-G001}  & {\rm Im}      & -14.93 & 22.12 & 504 &1271 & 0.85 & \dots & VII &  22 & 0.41 & 0.0 &\dots& 9.97 & 1.36& 0.104 \\ 
    {\rm UGCA 148}      & {\rm Im}      & -14.09 & 21.40 & 342 &1158 & 0.72 & 36    & VII &   6 &\dots & 1.5 &\dots& 0.34 &\dots& 0.038 \\ 
    {\rm UGCA 153}      & {\rm Sm/BCD}  & -14.21 & 23.32 & 937 &1427 & 0.42 & 52    & VII &  23 & 0.50 & 0.5 &\dots& 0.77 & 1.44& 0.187 \\ 
    {\rm NGC 2915}      & {\rm Sm}      & -16.61 & 21.26 & 673 &2290 & 0.56 & 62    & VII &  81 & 0.36 & 0.0 &\dots& 7.03 & 1.98& 0.091 \\ 
    {\rm UGCA 193}      & {\rm Sdm}     & -15.15 & 21.98 &1480 &4138 & 0.11 & 55    & VII &  23 & 0.56 & 0.5 &\dots& 2.36 & 1.24& 0.284 \\ 
    \noalign{\smallskip} 
    \hline
  \end{array} 
  $$  
  \begin{list}{}{}  \item[$^\dagger$]References to the published
    papers of our series on {\it Structure and stellar content of
    dwarf galaxies}: $I$ - Bremnes et al. (1998); $III$ - Bremnes 
    et al. (1999); $IV$ - Bremnes et al. (2000); $VI$ - Barazza et al. 
    (2001); $VII$ - Parodi et al. (2002).  
  \end{list}
\end{table*} 

Dwarf irregular galaxies exhibit peculiar morphologies that are
dominated by the flashy but seemingly irregular presence of
star-forming regions.  Unlike with more massive disk systems that
modulate star formation by spiral density waves, dwarf irregulars in
the field --- or non-interacting irregulars in general --- constitute
ideal testbeds for the study of genuine processes regulating local and
global star formation and consequently of galactic evolution. A review
addressing several key questions  concerning large-scale star
formation in irregular galaxies is given by Hunter (1997), and an
evaluation among  simple models for the onset of star formation in
irregulars  is provided by Hunter, Elmegreen, \& Baker (1998).

Important clues as to hidden constraints shaping the heterogeneous
appearance of dwarf irregular galaxies may emerge from detailed
investigations concerning the spatial distribution of star-forming
regions (e.g., Feitzinger \& Braunsfurth 1984). In recent years
several studies on the distribution of H$\;$II regions and of young,
compact star clusters in late-type spirals and in dwarf irregular
galaxies have appeared (Telles, Melnick, \& Terlevich 1997; Brosch,
Heller, \& Almoznino 1998; Elmegreen \& Salzer 1999; Heller et
al. 2000; Roye \& Hunter 2000; Billett, Hunter, \& Elmegreen
2002). Applying measures like concentration, asymmetry, and
fractional-luminosity indices these authors found the star forming
regions to be distributed rather randomly, with some tendency to
central concentrations particularly for star-bursting systems.

In this paper  we extend these previous studies by analysing the
distribution of bright spots or lumps in the $B$-band images of a
sample of 72 late-type (``irregular'') dwarf galaxies. The general
equivalence of bright lumps in H$\alpha$ and in broad band blue images
as tracers of star formation complexes can be appreciated by comparing
galaxy images filtered at the two corresponding wavelengths (e.g.,
Elmegreen \& Salzer 1999; Sparke \& Gallagher 2000, pp. 139 and
229). With our homogeneous and relatively large sample at hand we aim
at comparing three morphological indices with each other (as applied
to lumps within irregulars), none of which has been previously
reported in the form presented here or within our context. Indices
often serve as the quantitative counterparts to qualitative physical
concepts. In particular, lump spreading within a galaxy may be
described by {\it concentration indices} of different apertures; lump
clustering may be represented by the {\it correlation dimension} for
the two-dimensional lump distribution; finally, the {\it lumpiness}
(or flocculency) of a galaxy may be measured by means of some
fractional light index. Table 1 gives a summary of terms and indices
that will be more carefully introduced in the subsequent sections.

Comparing relations among morphological indices, we may deepen our
insights into the various mechanisms responsible for the morphology of
irregulars. The interstellar matter of dwarf irregulars with different
global properties may be different (e.g., metallicity, mean gas
density, turbulence, gravitational potential), implying differences in
the conditions for the formation of stars. Thus aspects concerning the
abundance and distribution of star-forming regions within the
galaxies, like clustering and star formation rates, may turn out to
vary correspondingly. We adress this issue by means of the clustering
parameter. Another goal of the paper is to contribute to the
discussion concerning the influence of shear due to differential
rotation on star formation in gas-rich late-type galaxies (Roye \&
Hunter 2000; Elmegreen, Palou\v{s}, Ehlerov\'{a} 2002).  Based on
cellular automata simulations we introduce a possible criterium to be
checked for in radial lump number distributions; we show that a few of
our galaxies indeed meet the criterium, but more research is needed
for conclusive results.

The paper is organised as follows. Section 2 gives an overview of the
galaxy sample used and presents a table with basic galaxy parameters
as well as the parameter values deduced in the subsequent
sections. Section 3 describes the adopted lump detection method and
introduces a first index, the lumpiness index. Section 4 presents the
radial number and  number density distributions for the bright lumps
of all the galaxies. A concentration index that is normalized
according to the galaxies being exponential-disk systems is introduced
and applied to the bright lump distribution. We then extensively
discuss the possible reason for a peculiarity seen in the lump number
distribution, namely the occurence and relative locations of major and
minor peaks. In Section 5 we determine the cluster or correlation
dimension for the two-dimensional lump patterns, relate it to galaxy
absolute magnitude, and deduce a model mean porosity that is linked to
the current  star formation rate. In Section 6 we check for relations
among the three indices introduced in the previous sections. We end
with a discussion and the conclusions in Section 7. Image processing
was performed throughout within the IRAF\footnote{IRAF is distributed
by the National Optical Astronomy Observatories;
http://iraf.noao.edu.}  package.

\begin{figure*}[th] 
\begin{center} 
 \includegraphics[width=85mm]{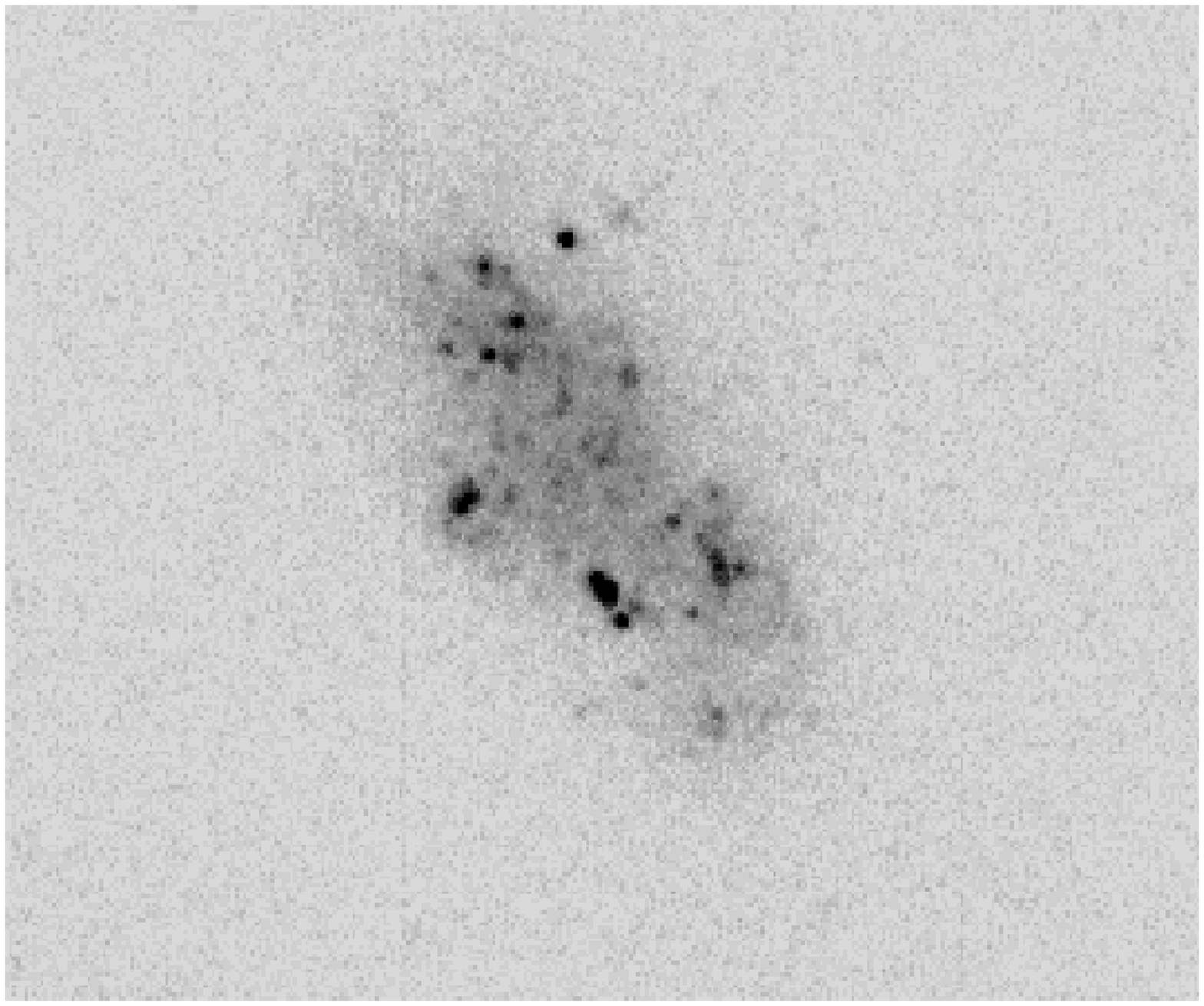} 
 \includegraphics[width=85mm]{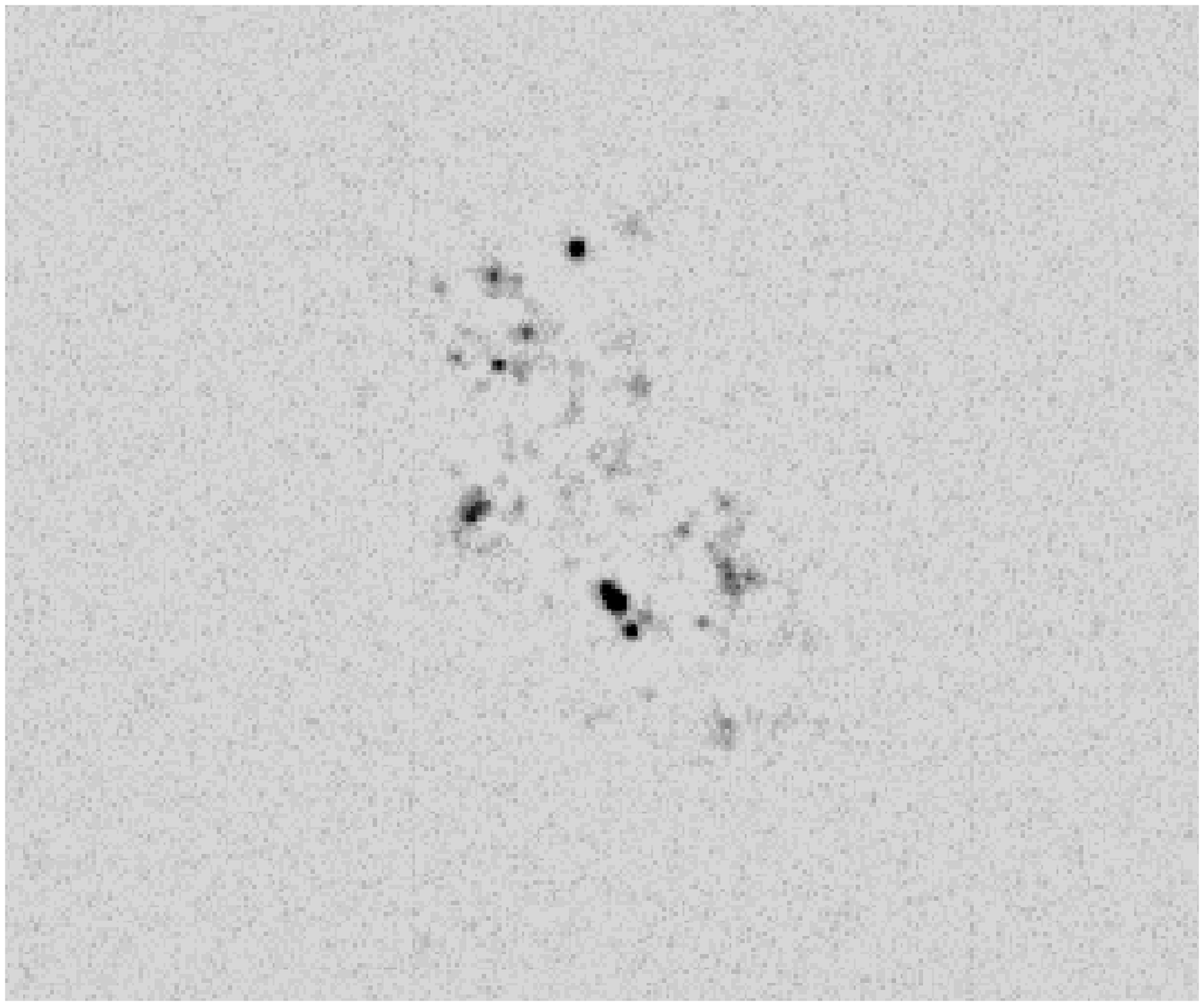} 
\end{center} 
\caption[]{ESO 473-G024. {\bf Left:} $B$-band image, taken at the 1.5-m 
Danish Telescope at La Silla, Chile. {\bf Right:} Residual image after 
subtracting from the original image its median filtered version.}
\label{eso473}
\end{figure*} 

\section{The sample}

The sample of 72 irregular dwarf galaxies is a compilation of Im, BCD,
and late-type S galaxies fainter than $M_B=-18$ mag --- dwarf
``irregulars'' for short --- used previously for an analysis of
exponential-disk model parameters relating photometric, kinematic, and
environmental properties (Parodi, Barazza, \& Binggeli 2002). The
galaxies are lying in the field or in groups within the nearby 10 Mpc
volume. They were imaged  with 1.2-m to 1.5-m telescopes with
resolutions of 0.39-0.77  arcsecs per pixel and under seeing
conditions varying between 0.8 and  4.0 arcsecs. Data reduction was
performed consistently by our team along the usual
prescriptions. $B$-band galaxy image galleries can be found in Parodi
et al. (2002) and the references given  therein (or at  the bottom of
Table 2 in the present paper).

In Table 2 we list basic galaxy data as well as parametric data as
deduced in the subsequent sections. The first nine columns are mostly
taken over from Table 1 in Parodi et al. (2002), while the seven other
columns correspond to results obtained in the present paper. The
columns read as follows:\\ 
Columns 1 and 2 give the galaxy name and the galaxy type;\\
columns 3 and 4 list the $B$-band absolute magnitude and --- from
fitting an exponential law to the observed surface brightness profile
--- the extrapolated central surface brightness, both corrected for
galactic extinction;\\
columns 5, 6, and 7 give the $B$-band disk scale length along the
semi-major axis (in parsecs), the radius of the 25th-mag/arcsec$^2$ 
isophote (in parsecs), and the axis ratio of the elliptical 
isophote, respectively;\\
column 8 lists the rotational velocities of the galaxies as far as
measured. The four entries for the southern hemisphere galaxies
ESO473-G024, IC2038, ESO490-G017, and ESO059-G001 are based on HIPASS
\footnote{The HI Parkes All-Sky Survey, or HIPASS, is a 21-cm HI
survey of the southern sky undertaken with a multibeam receiver on the
Parkes telescope in Australia. The one-dimensional spectral data for a
freely choosable position is available for downloading in a variety of 
different formats at {\rm
http://www.atnf.csiro.au/research/multibeam/release/.}} public
data, treated as in Parodi et al. (2002);\\
column 9 points to the five published papers of our series with
original and detailed  observational and photometric data on the
galaxies. Cf. the reference notes on the bottom of Table 2;\\
column 10 indicates the number of bright lumps found in residual 
galaxy images (Section 3);\\
columns 11, 12, and 13 give the scale length of the radial number
density  distribution of the bright lumps (Section 4.2), the radius of
the highest peak in the radial lump number distribution, and the peak
radius of a possible secondary, minor peak, respectively (Section
4.4) (all lengths in units of scale length $R_d$);\\
columns 14, 15, and 16 list the values for the following indices: the
normalized concentration index $CI$ measured using $R_{25}$ as the
outer aperture radius (Section 4.3), the reduced cluster (or
correlation) dimension $D$ (Section 5), and the lumpiness index \Chi 
(Sections 3 and 6), respectively.

\section{Lump detection and lumpiness index}

A residual image highlighting bright star-forming complexes ---
comprising H$\,$II regions and young clusters --- was constructed by
subtracting a smoothed version from the original $B$-band image for
each galaxy. To obtain the former image a convenient method is to
median filter the original image with a sliding square window. Relying
instead on adaptive filtering techniques (i.e., on the IRAF task
ADAPTIVE) did not improve on the results, thus we kept applying the
simple median filtering method. For the window $w$ some characteristic
metric size should be chosen: we adopted $w = 0.2 R_{\rm eff}$, with
$R_{\rm eff}$ being the effective radius of the galaxy.  An example of
images processed this way can be seen in Fig. \ref{eso473}. The galaxy
shown is ESO 473-G024, with the left frame containing the original
image, and the right frame presenting the residual image. From these
latter, high-spatial frequency images we extract the following
information:

(i) we detect and tabulate the locations of the bright lumps, thus
constructing a data base for the distribution analysis below.
Detection of bright lumps was done automatically and thus
consistently for all galaxies. First, the sky-subtracted residual
image is cleaned around and, concerning obvious foreground stars,
within the galaxy. The few foreground stars that probably went
undetected should have no potential to fake the  statistical outcomes
of this paper. Second, an appropriate point spread function (PSF) was
looked for: a routine checks for the maximum pixel value corresponding
to the brightest lump and deter\-mines its PSF. It may happen that the
brightest lump takes part in a compact cluster of lumps; thus if the
PSF's full-width-at-half-maximum was more than 10 pixels the routine
assumed blending and searched for a smaller lump with a narrower
PSF. Similarly, if the PSF was very peaky, an overseen foreground star
was assumed and a broader PSF was applied for the lump search. Third,
this PSF is then used by the IRAF task DAOFIND to search for all other
lumps above a detection threshold of 3 sky sigmas. In ESO 473-G024,
for example, 22 bright lumps were thus counted. Some galaxies are too
fuzzy-looking to have any lumps detected. Finally, lump  coordinates
are stored in physical units with the center and the major axes of the
25th-mag/arcsec$^2$ isophotal ellipse providing the origin and the
axes of the coordinate system; {\it assuming axisymmetry, the lump
coordinates will be deprojected to zero ellipticity for all
applications}. Note that we do not estimate lump luminosities in
order to compare lump brightnesses or to provide luminosity
functions; for blue luminosity  functions of star-forming complexes in
spiral and irregular galaxies see Elmegreen \& Salzer (1999).
 
(ii) We determine the galaxy's lumpiness index \Chi. Lumpiness (or
flocculency) of a galaxy may be quantified by the high spatial
frequency power \Chi, pioneered by Isserstedt \& Schindler (1986) and
recently applied by Elmegreen \& Salzer (1999) to spiral and irregular
galaxies and by Takamiya (1999) to HDF-N galaxies with the aim of
having a galaxy structural parameter at hand that is related to the
current star formation rate. Following the notation of Takamiya (1999)
it is defined as the ratio of the flux from the bright lumps, $f_l$,
and the total flux $f_g$ of the galaxy, thus
\[ \Chi = \frac{f_{\large l}}{f_g}\,. \]
By measureing the total fluxes of the residual and the original images
for each galaxy, we obtained the \Chi\/ values listed in Table 2. The
lumpiness index will be applied below in Section 6.

\section{Radial distributions of detected bright lumps}

\subsection{Number distribution}

\begin{figure*}[th] 
\begin{center}
  \includegraphics[width=130mm,height=180mm]{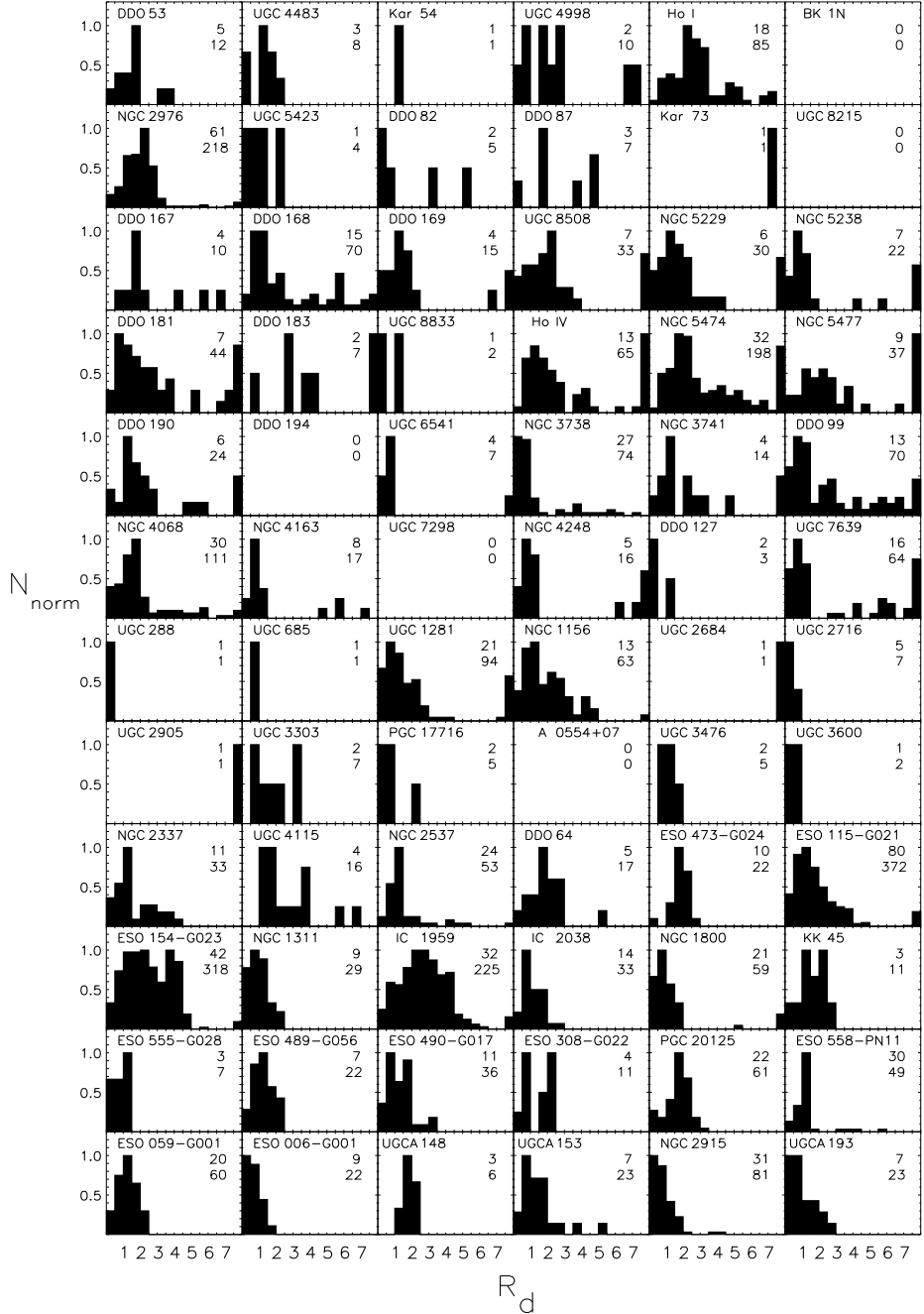} 
\end{center} 
\caption[]{Radial number distribution of bright lumps in 72 dwarf
irregular galaxies, based on B-band images. Bin width is 0.5 $R_d$;
the rightmost bin sums all counts beyond 7.5 $R_d$. For each galaxy
the counts are normalized by the highest bin value. In each panel the
galaxy name, the highest bin value, and the total number of counts are
given.}\label{radial_distr_indiv}
\end{figure*} 
\begin{figure*}[th] 
\begin{center} 
  \includegraphics[width=130mm,height=180mm]{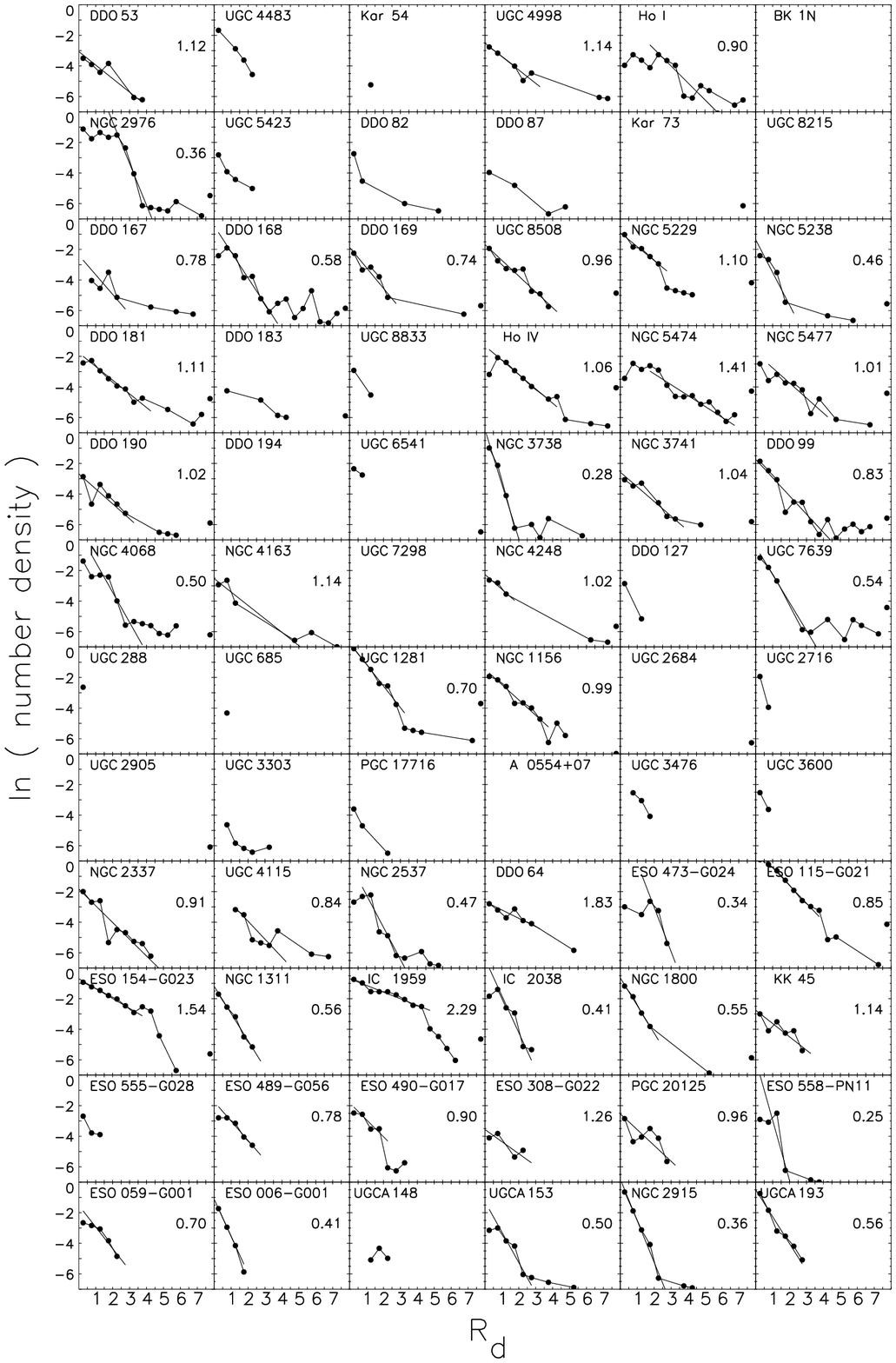} 
\end{center} 
\caption[]{Radial number density distribution of bright lumps in 72
dwarf irregular galaxies, deduced from the number distribution shown
in  Fig. \ref{radial_distr_indiv}. In each panel the galaxy name and
the distribution's approximate exponential-fit scale length are given
(in units of $B$-band  scale lengths;  evaluated only if there is a
total of at least 10 detected lumps, and represented by straight
solid lines).}
\label{radial_distr_distri}
\end{figure*} 

In Fig. \ref{radial_distr_indiv} we show for all galaxies of the
sample the binned radial number distributions of the detected bright
lumps. Histogram bins correspond to concentric elliptical annuli with
semi-major axes successively growing by half a scale length $R_d$ and
shown out to 8 $R_d$; the last bin included in the panels comprises
all the detected lumps at radii larger than 7.5 $R_d$. The number of
counts per bin, $N_{\rm norm}$, is normalized by the largest bin value
within 7.5 $R_d$; the largest bin value as well as the total number of
counts are printed within each panel. This kind of normalization was
imposed to minimize the effects of different resolutions and 
seeing conditions for the following inquiry.

\begin{figure}[t] 
\begin{center} 
  \includegraphics[width=90mm,height=60mm]{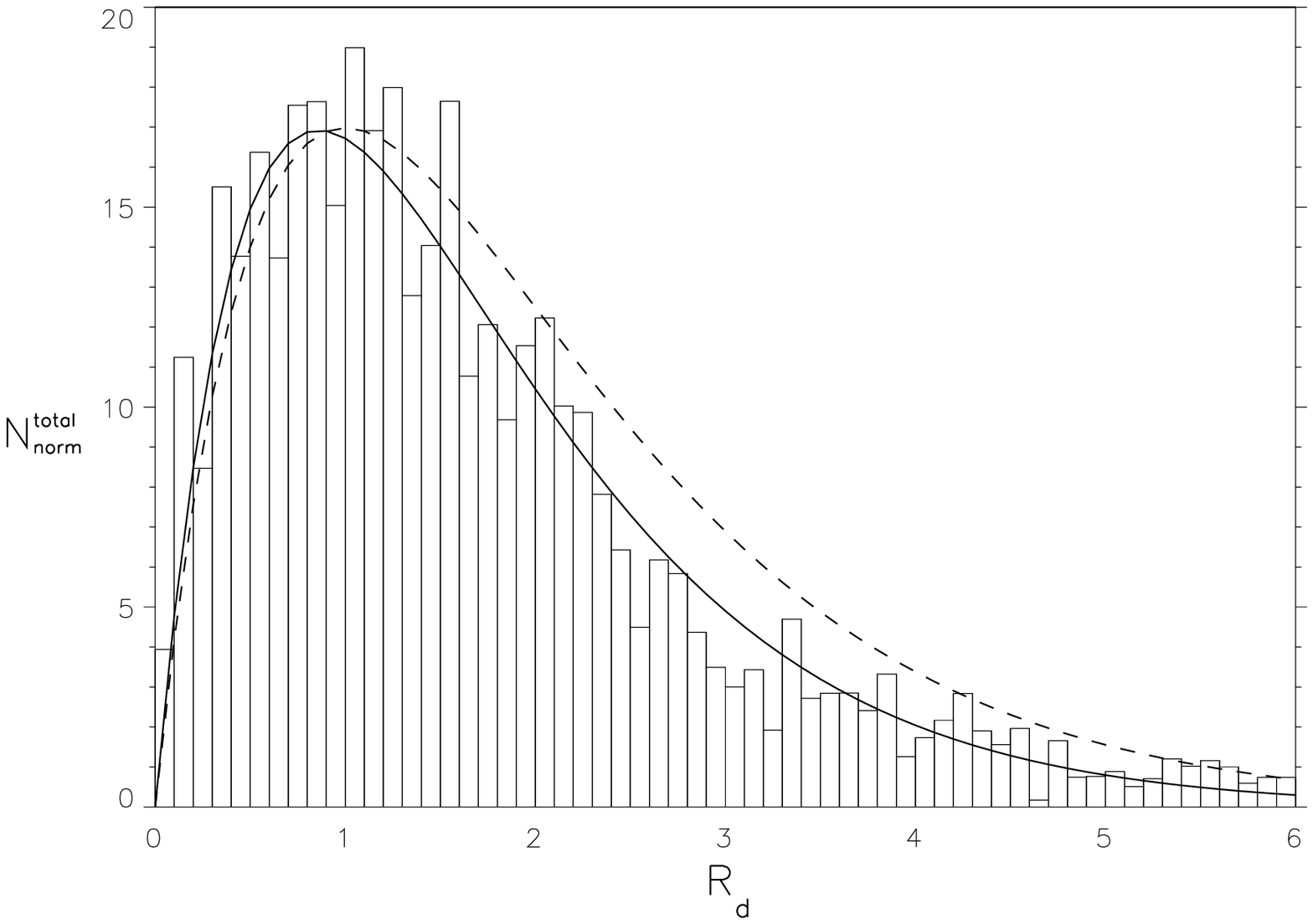}
\end{center} 
\caption[]{Total radial number distribution of bright lumps  in 72
dwarf irregular galaxies. Binning in units of a tenth of a scale
length; the contributions from each galaxy are normalized by the value
of its highest-value bin (cf. Fig. \ref{radial_distr_indiv}). The
dashed line follows a distribution $R\,{\rm exp}(-R/R_d)$ as it
is  expected from the continuum light exponential disks. A better
match  is provided for a distribution with a mean lump scale
length  of $R_l=0.86 R_d$, plotted as solid line.}
\label{radial_distr_cumul}
\end{figure}

The accumulated radial number distribution $N_{\rm norm}^{\rm total}$
for the lumps of all the galaxies, i.e. summing up all the profiles of
Fig. \ref{radial_distr_indiv}, is shown as histogram in
Fig. \ref{radial_distr_cumul}. The bright-lump distribution of many
galaxies thrown together is represented by a radius-weighted
exponential distribution that is indicated by the solid line obeying
$N(R)\propto R\,exp(-R/R_l)$ with $R_l=0.86 R_d$ (cf. Section
4.2). This basically reflects the exponential light profiles of dwarf
irregular galaxies in general with, however, a slightly shorter scale
length than is seen for the $B$-band continuum light (dashed
line). Note that bright lumps or  star-forming complexes not only are
found way out to large radii, but that they constitute a nice tail in
the radial number distribution out to at least six optical scale
lengths, a point we come back to in the discussions of  Section 7.

The similar exponential structure for two components of the disk is
comparable to H$\;$II region distributions in other types of
exponential-disk galaxies. In intermediate-type spirals (Athanassoula
et al. 1993) and in irregular galaxies (Hunter et al. 1998) the
azimuthally averaged radial distribution of H$\;$II regions follows
the stellar light distribution as well. In our dwarf irregular
galaxies bright lumps are exhibiting this same behaviour, indicating
them to be representative for the distribution of H$\;$II regions, as
expected. We can put this statement on a still firmer basis tracking
down also the radial number density distribution of the bright lumps.

\subsection{Number density distribution}

Radial number density distributions for the bright lumps are obtained
by dividing the number of counts in a given bin by the surface of the
corresponding elliptical annulus (cf. Hodge 1969, Athanassoula et
al. 1993). The bright-lump number density profiles for our galaxies
are shown in Fig. \ref{radial_distr_distri}. Solid lines represent
exponential fits for galaxies with a total of at least 10 detected
lumps; their scale lengths $R_l$ are given in each panel in units of
$B$-band continuum light scale lengths $R_{d}$ and are listed in
Table 2. For the whole sample we find a mean of
\begin{equation} <R_{\/l}/R_d> = 0.86\pm 0.06, \;\;\sigma =
0.41\;.
\end{equation} Thus, while the number density distributions of bright 
lumps  are --- at least partially --- also exponential, their slopes
are on average steeper than those for the density distributions of the
underlying stellar light. We note that this very same quantitative
behaviour was also observed for the number density distribution of
H$\;$II regions in intermediate-type spiral galaxies: Athanassoula et
al. (1993) found $<R_{H\,II}/R_d> = 0.8$, $\sigma = 0.4$.\\

Concluding this subsection we state that dwarf irregular galaxies show
azimuthally summed-up bright-lump profiles that are quantitatively
comparable to those of H$\;$II regions in exponential-disk
systems. Thus, as expected, bright star-forming complexes largely
represent H$\;$II regions. In particular, the scale lengths for
H$\;$II regions, for bright lumps on $B$-band images, and for $B$- and
$R$-band continuum light images (Parodi et al. 2002) obey on average
the ratio equation $R_{HII}:R_l:R_d(B):R_d(R)\approx0.8:0.9:1.0:1.1$,
i.e. the older the underlying population the larger the scale lengths.
This general trend in star-forming dwarfs has been observed before,
and we discuss some implications in Section 7.
\begin{figure}[th] 
\begin{center} 
 \includegraphics[width=85mm]{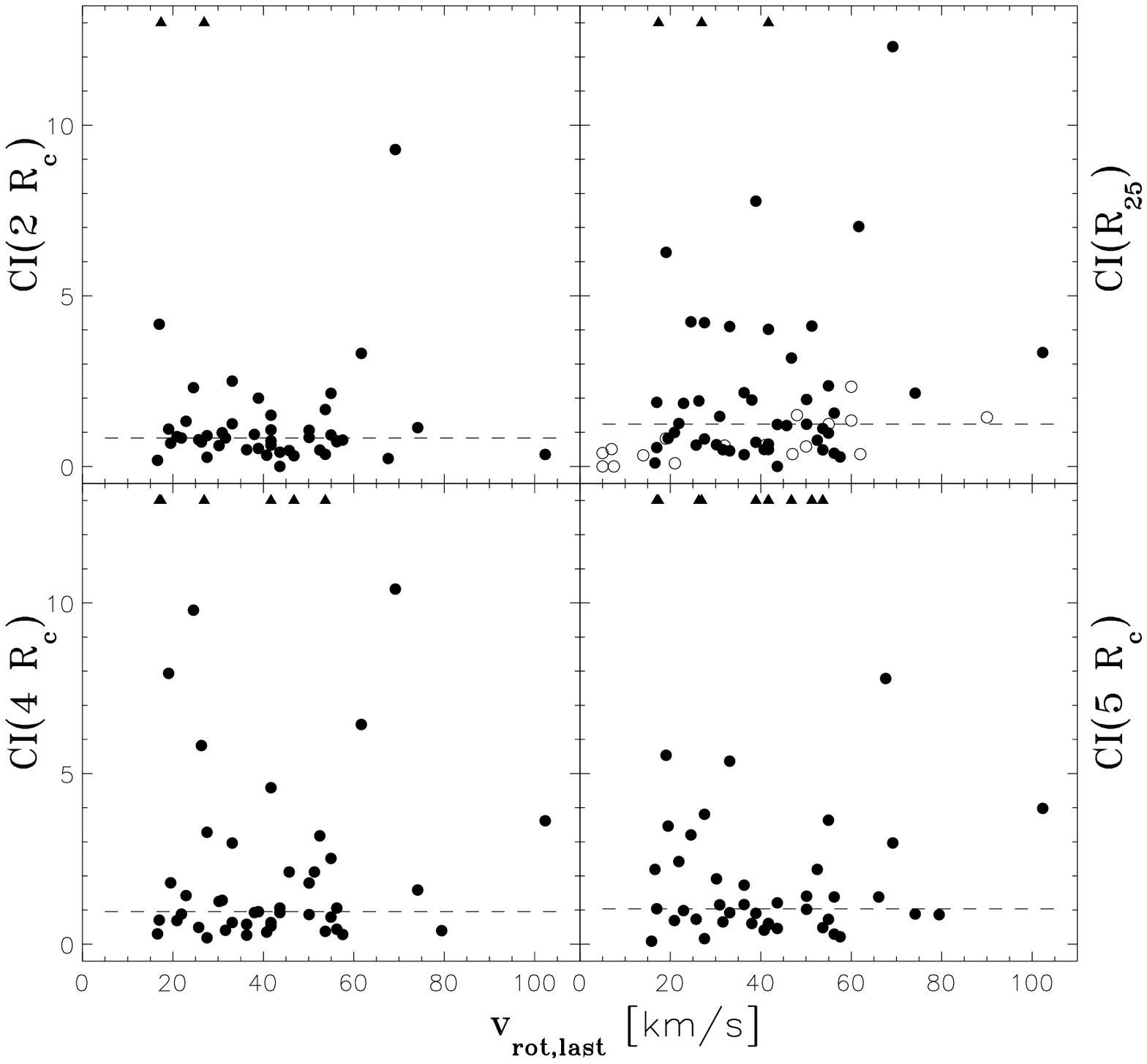} 
\end{center} 
\caption[]{Normalized bright-lump concentration indices $CI$
vs. galaxy rotation velocity. Pure exponential-disk distributions
correspond to $CI=1$. Filled circles represent galaxies of our sample,
triangles at panel upper boundaries stand for infinite $CI$ values
resulting from galaxies with no lumps detected in the outer annulus,
and open symbols is data from Roye \& Hunter (2000).}\label{CI_vrot}
\end{figure} 

\subsection{Concentration index}

The concentration index $CI$ of a galaxy is a convenient parameter to
quantify galaxy morphology of low- and high-redshift galaxies. It
compares the light content for different radial intervals. Various
definitions have been used, none of which takes into account the
exponential-disk constraint. It is, however, a trivial observation
that in scale length-versus-luminosity diagrams the scatter in scale
length around the mean relation correlates with the concentration
index. To have a scale length-independent concentration index we will
explicitely factor in this underlying disk feature. Following Heller
et al. (2000) for the sake of comparision, the concentration index
$CI(R)$ is taken as the ratio of the flux or, in our case, the number
of complexes within an elliptical aperture of semi-major radius $R/2$,
i.e. from the {\it inner part} of the galaxy, to the flux or number of
complexes from its {\it outer annulus} with inner and outer semi-major
radii of $R/2$ and $R$, respectively. Opposite to Heller et al. (2000)
--- who suspect linear radial distributions of the (flux from) star
forming regions --- we do not bring the two numbers to an equal-area
basis by dividing the outer number of lumps by a factor of
three. Instead, we want to relate the measured lump $CI$ to the
corresponding one for an assumed underlying radial number distribution
$N(R)\propto R\,exp(-R/R_d)$ (or, as it is more adequate for lumps,
$\propto R\,exp(-R/R_l)$; see below). We thus {\it normalize} our galaxy
concentration index with $CI_0(R) = \int_0^{R/2} N(r) dr /
\int_{R/2}^R N(r) dr$. Expressing the total radius $R$ in terms of the
disk scale length $R_d$, $R=x\,R_d$, one finds
\begin{equation}
CI_0(x) = \frac{\left[ 1-(1+x/2)e^{-x/2} \right]} {\left[
(1+x/2)e^{-x/2}-(1+x)e^{-x} \right]}\,.
\end{equation}
Thus our normalization constant $CI_0(x)$ depends on the chosen outer
radius $R$, or $x$, respectively, used to determine the concentration
index. On a statistical footing the 72 dwarf irregular galaxies of our
sample share the mean ratios $R_d : R_{eff} : R_{25} : R_{995}$
$\approx$ $1 : 1.5 : 3 : 5.1$ in the continuum $B$-band light, where
$R_{eff}$ is the effective radius, $R_{25}$ is the 25th-mag/arcsec$^2$
isophotal radius, and $R_{995}$ is the radius equivalent to an
aperture with 99.5\% of the total flux. Theoretically, pure
exponential-light decays yield $R_d : R_{eff} : R_{25} : R_{995}$ $=$
$1 : 1.7 : 4.1 : 7.4$, where the value for $R_{25}$ was interpolated
using the observed mean ratios given above. The fact that our galaxies
exhibit on average smaller radii is a consequence of the
exponential-disk description being only an approximation; in
particular, there is a large fraction of bright galaxies with central
light cusps. The appropriate normalization constants corresponding to
the above radii are then $CI_0(1)$=0.52, $CI_0(1.5)$=0.64,
$CI_0(3)$=1.23, and $CI_0(5.1)$=3.01. 

{\it With this normalization we expect the concentration indices to be
$CI\approx1$}. Deviations from this canonical value provide
information on the actual shapes of the profiles.  For example, the
median values of the concentration indices obtained by Heller et
al. (2000) for their dwarf irregular (BCD + LSB) galaxy sample are ---
corrected to our normalization by adopting their limit  $R=R_{25}$ and
thus applying $CI_0=1.23$ --- (4.23/3)/1.23 = 1.15 and
(3.43/3)/1.23=0.93 for the H$\alpha$-flux and the continuum images,
respectively. Thus BCD and LSB galaxies are rather well represented by
exponential light profiles on large parts. However, the values for the
BCDs typically lie above and those for the  LSBs below these median
values, reflecting the fact that the former galaxies are more actively
star forming in the center regions than the latter. Returning to our
sample, we arrive at values  $CI=$0.94, 1.26, 1.32, 1.73 for outer
radii $R=$2, 3, 4, 5 $R_d$,  respectively. These values typically
being larger than one and even increasing with larger outer radii is
due to  the normalization used so far that was based on the continuum
light  scale length $R_d$.

However, if the lump scale length $R_l$ is used for the normalization
instead of $R_d$, one indeed recovers $CI\approx1$. This is shown in
Fig.$\,$5 where we have plotted the normalized concentration indices
$CI(x)$ of our galaxies for various aperture radii $R=x R_l$ against
their rotational velocities, now adopting the mean scale length
$R_l=0.86 R_d$ found for the lumps in Section 4.2. Actually, one
would prefer to adopt for each galaxy its particular lump scale
length, but given the uncertainties in determining them, we are
content with the mean value given in equation (1). Only galaxies with
at least five detected lumps are included, leaving about 50
galaxies. Infinite values result for $CI$s in the case of no
outer-annuli lump detections; in the plots they are included  as
triangle symbols with values fixed at $CI=13$. Data for the upper
right panel is listed in Table 2. The panel's median $CI$ values,
plotted as dashed lines and ignoring the $CI$=$\infty$ cases, are
0.83, 1.24, 0.95, and 1.04.  Lying all in the vicinity of one, this is
consistent with the annulus-integrated exponential distribution for
the summed lump number distribution seen in Fig. 4 out to large radii.

Thus, applying the analytic tool of the concentration index, we have
again demonstrated that {\it the radial distribution of star forming
regions is non-linear but follows an annulus-integrated exponential
distribution}. This implies a non-uniform random spread of the star
forming regions throughout the disk, which explains the discrepancy
found by Heller et al. (2000) between the $CI_{H\alpha}$ values for
actual galaxies and the lower ones for simulated galaxies with random
star formation region positions. Roye \& Hunter (2000) pointed out an
increased scatter of concentration indices for faster rotating
galaxies of their sample. In the upper right panel, data from Roye \&
Hunter (2000), adopted to our normalization using $CI_0(3.5)=1.53$
(assuming $R_{25}$=3$R_d$=3.5$R_l$) and with a median $CI$ value of
only  0.58, are plotted, too; however, we no longer see this effect in
any  panel with our larger sample.

\subsection{Peak number distribution: shear-enhanced star formation at work?}

\begin{figure*}[th] 
\begin{center} 
   \includegraphics[width=8.9cm]{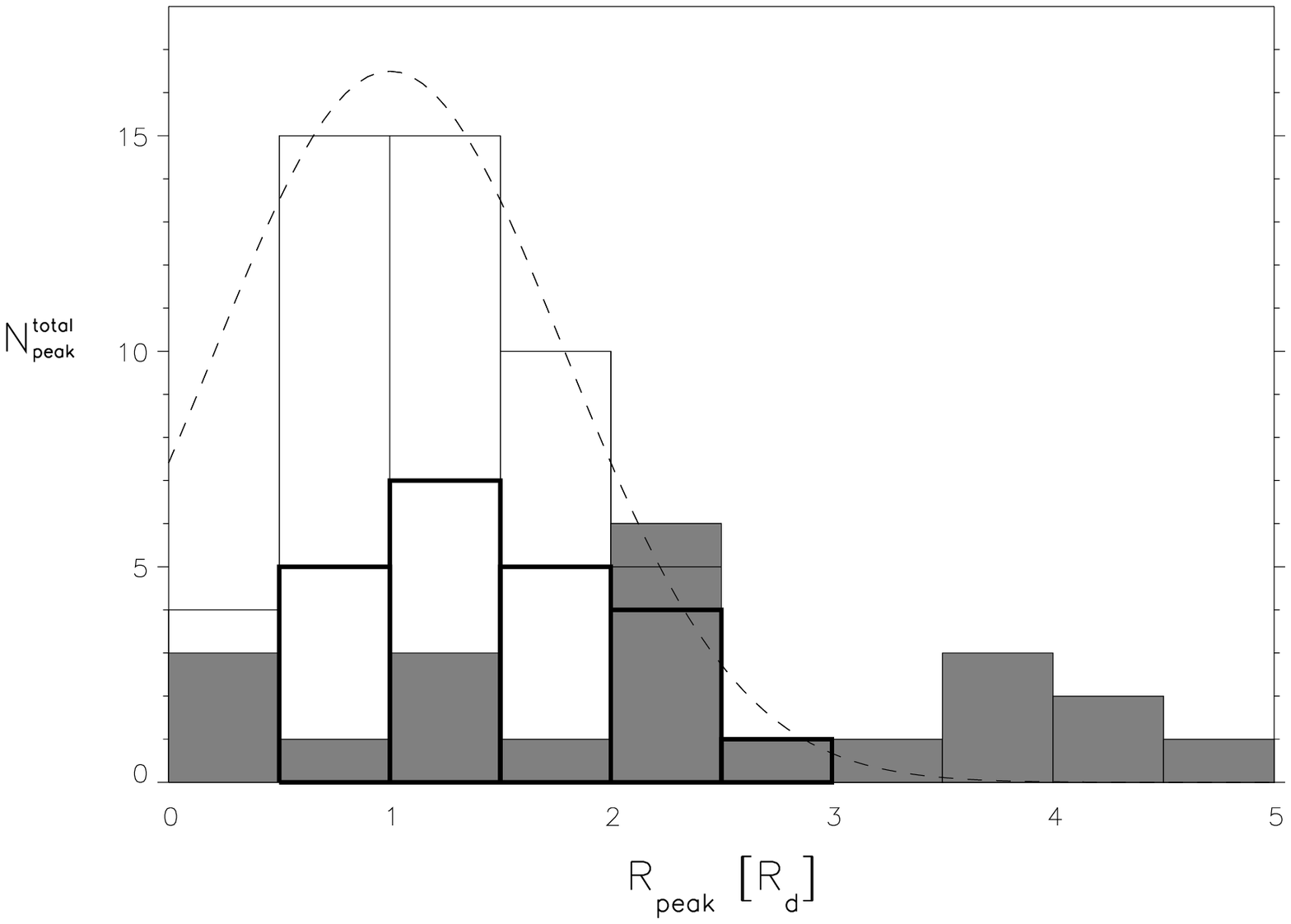}
   \includegraphics[width=8.9cm]{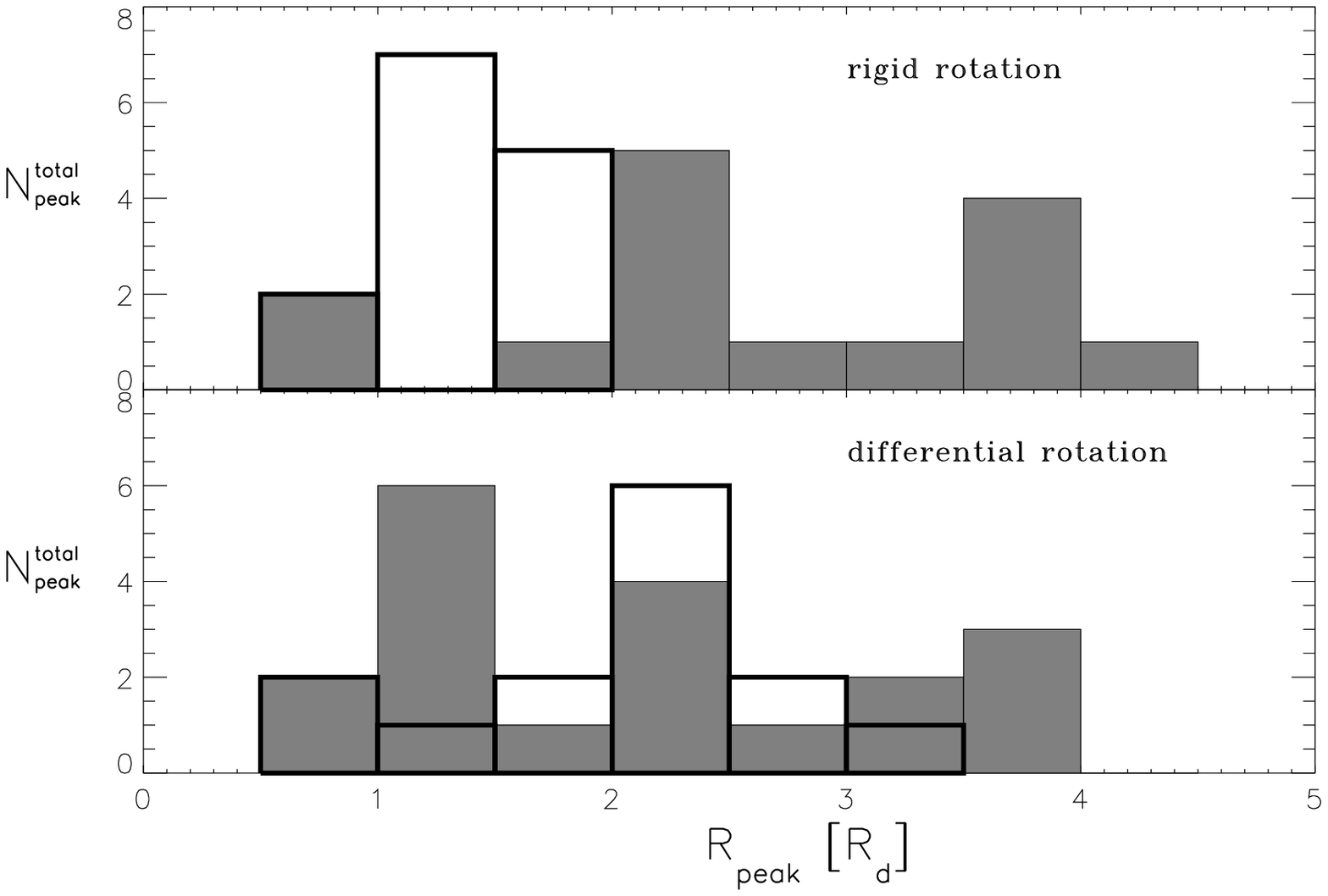}\\
\end{center} 
\caption{{\bf Left:} Peak number distribution of bright lumps in
dwarf irregular galaxies, deduced from the number distributions shown
in  Fig. \ref{radial_distr_indiv}; included are only galaxies with
peaks corresponding to at least three counts. The histogram for all
the primary (i.e. highest) peaks is shown as bars bordered with thin
lines; the dashed line is a Gaussian eye fit forced to peak at one
scale length. The histogram for the primary peaks of those galaxies
exhibiting secondary peaks as well is overplotted with thick lines;
the shaded bars indicate the distribution of the corresponding
secondary peaks. Note the pronounced signals at the center and in
particular at two scale lengths.  {\bf Right:}  Peak number
distribution for 20 simulated galaxies, once for galaxies with
solid-body rotation (upper panel) and once for galaxies with a
transition to strong differential rotation at two scale lengths (lower
panel). The simulated galaxies are generated by means of a stochastic
self-propagating star formation (SSPSF) disk model. As in the figure
to the left, thick-lined bars represent primary peaks whereas shaded
bars inform on secondary peaks. Note the reversed peak distribution.}
\label{radial_distr_peaks}
\end{figure*} 

For the sample as a whole the radial locations of bright lumps mirror
the exponential intensity distribution of the underlying
population. This is, however, only a rule-of-thumb. Individual
galaxies may exhibit strong deviations from this mean statistical
behaviour (cf. Fig. \ref{radial_distr_indiv}). For example, the
primary peak in the radial number distribution (i.e. the bin with the
highest value) appears not around one scale length, but is shifted to
smaller or higher radii. The histogram for the radial distributions of
primary peaks only is shown in Fig. \ref{radial_distr_peaks}, left
(thin-lined bars). Only galaxies with peaks corresponding to at least
three counts are included. The expected maximum of occurences of
primary peaks around one scale length is clearly recovered; a Gaussian
with a mean at one scale length and a standard deviation of
1.4$\;R_{\rm d}$ is overplotted as the dashed line. However, another
particular feature of individual lump number histograms is the
frequent presence of a secondary peak that is lower than the main peak
(instead of monotonicly smaller bin heights on both sides of the main
peak);  this is the case for about a third of our galaxies
(cf. Fig. \ref{radial_distr_indiv}). The radii of primary peaks (with
at least three counts) and secondary peaks are listed in Table 2; in a
few cases of equal height peaks we refered to 0.1 $R_d$-bin width
number distributions to decide which of the peaks is the primary or
the secondary one. {\it While the primary peaks of those galaxies
exhibiting a second, minor peak as well  are crowded around one scale
length} (thick-lined bars), {\it the corresponding distribution of
secondary peaks reveals a pronounced maximum at about two scale
lengths} (shaded bars).

Is this excess of bright stellar complexes at radii larger than about
two scale lengths a statistical fluctuation or is it a manifestation
of some underlying mechanism? A candidate mechanism that deserves
closer inspection is shearing due to differential rotation within the
outer part of the disk. We thus discuss the physical plausibility for
the influence of shearing on the generation of star-forming complexes
within dwarf galaxies. Two questions will be adressed: First, {\it is}
shearing in dwarf galaxies a viable mechanism? And second, can it
account for the observed peak distribution?

A compilation of 20 high quality dwarf galaxy rotation curves by
Swaters (2001) shows them to look much like those of spiral galaxies,
with rotation curves rising steeply in the inner parts and flattening
in the outer parts. In particular, most {\it dwarf galaxy rotation
curves start to flatten around two disk scale lengths}, and no dwarf
galaxy shows solid-body rotation beyond three disk scale lengths
anymore. Concerning our observed occurence of minor peaks in the
bright lump distribution, starting at and being most pronounced at
about two scale lengths as well, we may wonder whether it is related
to the transition from solid-body to differential
rotation. Affirmative signals arrive both from theory and
simulations. (i) Larson (1983) suggested that the SF rate increases
with higher shear rate via the ``swing amplifier'' mechanism: citing
Toomre he points out that shear itself contributes strongly to the
growth of gravitational instabilities, leading to gas density
enhancements and subsequent star formation. For dIs, however, lacking
spiral-density waves, swing amplification may seem an inappropriate
mechanism to rely on. (ii) Alternatively, in their review Seiden \&
Schulman (1990, p. 40) remind that in models for stochastic
self-propagating star formation (SSPSF) shearing increases the density
of star-forming regions: gas-rich, potential star-forming regions are
transported to and mixed with  former star-forming regions, giving
space for new star formation. (iii) Additionally, while it has been
questioned whether shear may cause any visible effect at all given
dwarf galaxies being rather slowly rotating systems (Hunter et
al. 1998), it becomes more and more evident that some irregular
galaxies like the Large Magellanic Cloud or NGC4449 possess regular,
large-scale magnetic fields (e.g., Otmianowska-Mazur et al. 2000), and
thus it similarly becomes feasible that the magneto-rotational
instability (e.g., Balbus \& Hawley 1998) comes into play, effectively
strengthening the effects of shear. Bearing in mind this possibility,
we nevertheless restrain the discussion in the following on the second
of these scenarios only.

Self-propagating star formation is observed with many galactic as well
within many extragalactic objects. It is thought of as a locally
important SF triggering mechanism in all types of galaxies. For
example, modulated by density waves, long-lived spiral arms may be
formed in bright disk galaxies (Smith, Elmegreen, \& Elmegreen 1984);
the surface filling factor of bubbles and the locations of molecular
rings in observed disk galaxies can be quantitatively explained by
SSPSF (Palou\v{s}, Tenorio-Tagle, \& Franco 1994); age gradients in
star-bursting galaxies can be accounted for by means of triggered star
formation (Thuan, Izotov, \& Foltz 1999; Harris \& Zaritsky 1999); and
last but not least, the general burst characteristics of compact and
irregular dwarf galaxies is long known to partially be understood by
means of  SSPSF (Gerola, Seiden, \& Schulman 1980; Comins 1984).

Relying ourselves on a two-dimensional SSPSF model, we numerically
tested the hypothesis that the onset of shear-induced star formation
around the turnover radius may leave its imprint in an overabundance
of SF regions or of bright stellar complexes beyond two scale
lengths. In the Appendix we describe the particular implementation. A
general finding of our simulations is that the inclusion of shear
(i.e. rotation) allows for about five to ten percent more star forming
cells, the exact value depending on the particular parameters
used. Being mainly interested in the azimuthally summed-up radial
distribution of lumps under different rotational conditions, we
compare simulation runs with and without  a transition to a flat
rotation curve. In the top panel of Fig.  \ref{radial_distr_peaks},
right, a typical outcome for a simulation of 20 galaxies with rigidly
rotating disks is plotted. The highest peaks are found to be located
around one scale length, while the secondary peaks show occurences at
many radii but with a preference for locations around two scale
lengths. This coincides with our observed peak distributions. For
comparision, in the bottom panel of Fig. \ref{radial_distr_peaks},
right, a representative peak distribution for a simulation run of 20
galaxies that exhibit a continuous transition from solid-body to
(strong) differential rotation at two scale lengths is
shown. Interestingly, the primary peaks now occur preferentially at
around two scale lengths indicating a strong influence of shear on
star formation around the turnover radius. While this is not the
general picture observed with our sample, some of the  {\it brighter}
galaxies actually do match this  pattern: IC 1959, ESO 154-G023, Ho I,
NGC 5477.

We thus conclude that the observed pattern of primary peaks at one
scale length manifests the underlying exponential-disk systems, and
that {\it the frequent occurence of secondary peaks at about two scale
lengths is not necessarily related to the onset of strong shear in
rotating disks. As the simulations show, it is however consistent with
the idea of triggered star formation based on a stochastic
self-regulation scenario}. Some of the larger galaxies are exhibiting
pronounced primary peaks at two scale lengths but show minor peaks at
one scale length; with these galaxies we  may be directly witnessing
shear-induced star  formation. The possible role of bars will be 
reflected in Section 7.

\section{Clustering properties of bright lumps: cluster dimensions on 
scales of a few 100 pc}

\begin{figure*}[th] 
\begin{center} 
 \includegraphics[width=85mm]{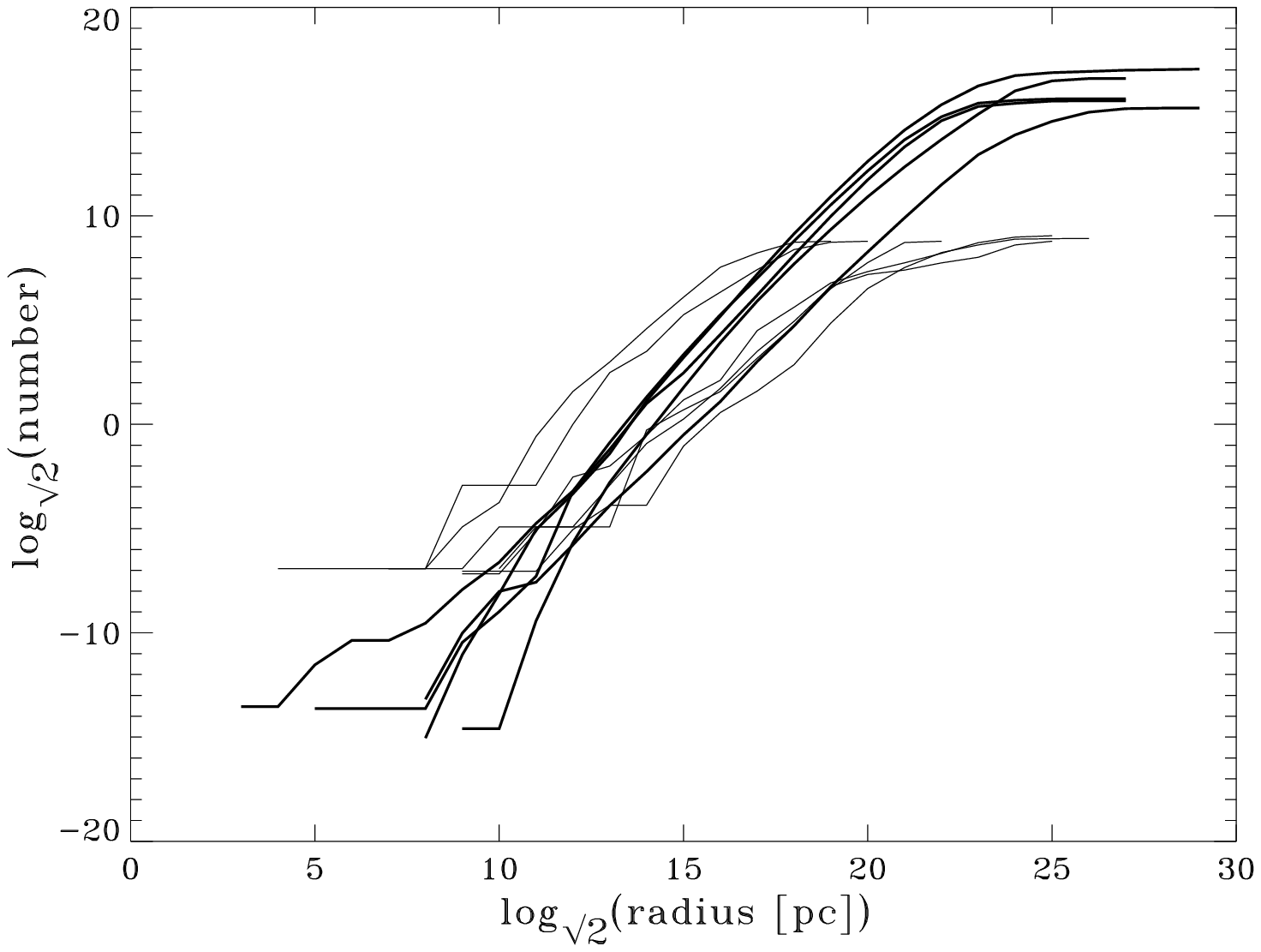} 
 \includegraphics[width=85mm]{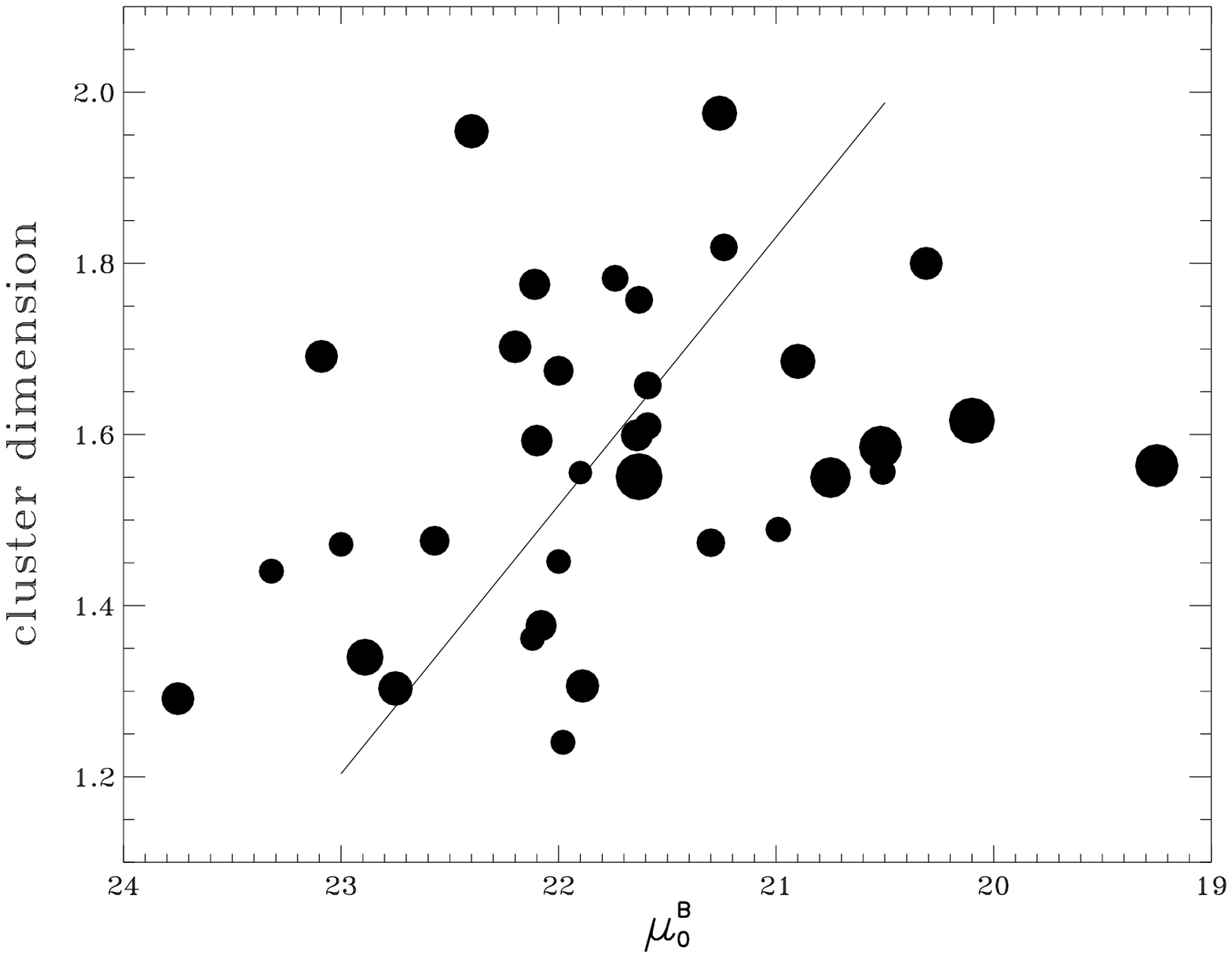} 
\end{center} 
\caption[]{{\bf Left:} Cumulative number vs. aperture radius for the
detected lumps within selected galaxies. Logarithms are given  to
base $\sqrt{2}$. The slope of the straight parts gives the cluster
dimension. Fat lines represent large-scale length (and high-lump
number) galaxies, thin lines short-scale length (and low-lump number)
galaxies. {\bf Right:} Reduced bright-lump cluster dimension
vs. extrapolated central surface brightness. The symbol size indicates
the number of lumps used for  the determination of the observed
(non-reduced) cluster dimension. The line corresponds to a bisector
fit  to the data (equation given in the text).}\label{clusterdim}
\end{figure*} 

Star-forming complexes in dwarf galaxies form non-random point
patterns also in a sense different from that discussed in Section
4.3. Their positions correlate according to a {\it self-similar
(fractal) arrangement}. In this Section we substantiate this claim
studying an index devoted to spatial statistics, namely the
correlation or clustering dimension, as applied to two-dimensional
bright-lump distributions.

As observed by Elmegreen \& Elmegreen (2001), the distribution of
bright-lump center positions on a kiloparsec scale in spiral and
irregular galaxies obey a power-law behaviour similar to the fractal
structure of the interstellar gas with fractal dimension $D_3=2.3$.
Thus the center positions of star-forming aggregates within {\it
isolated areas} of large galaxies are fractal. Here we address the
question whether star-forming complexes that are scattered over the
{\it entire disks} of dwarf galaxies are non-randomly distributed as
well. We restrict our inquiry to the dwarfs of our sample that exhibit
more than 20 bright lumps. Given our photometry with image scales of
typically well above 10 parsecs/pixel and seeing conditions of a few
pixels we expect to only dissolve structures larger than about 100
parsecs. Thus small-scale clustering and the accompagning blending
effects (Elmegreen \& Elmegreen 2001) are of no concern to our
study. To quantify the spatial clustering of the position patterns we
adopt the cumulative distance method (Hastings \& Sugihara 1996): a
power law relationship $N(r) \propto r^D$ is assumed for the
cumulative number of points $N(r)$ within a distance $r$ around each
point. If the distribution is (at least partially) self-similar this
will be manifested in a $log(N)$-$log(r)$ diagram as a straight line
with slope $D$, called the cluster (or correlation) dimension, \[ D=
\frac{d\,\log N(r)}{d\,\log r}.\] The more highly clustered the points
(at all relevant scales), the lower the cluster dimension. For a
random or Poissonian distribution of points on a two-dimensional plane
one has $D\approx2$, {\it independent} of the number of points
involved, which only governs the error estimate.  The graphs for six
observed galaxies with 20-30 lumps and for five galaxies with about
200-300 lumps are shown in Fig.\/7, left, plotted with  thin and
thick lines, respectively. For both groups the relevant scaling range,
i.e. the straight part of the curve, lies between about 100
($\approx\sqrt{2}^{13}$) and 1000 ($\approx\sqrt{2}^{20}$)
parsecs. The galaxies with lower lump numbers exhibit  smaller
cluster dimensions ($D\approx1.5$) than the galaxies with many
detected lumps ($D\approx1.9$). However, plotting $D$ versus
$N_{lumps}$ for {\it} all our data (not shown), {\it no} clear
relation between the two quantities is seen anymore. There
nevertheless {\it is} a hidden dependence between the two variables:
it emerges from the non-uniform  distribution of lumps in
exponential-disk systems (as discussed in Section 4), and it is to be
corrected for.  We do so by, first, simulating point patterns with
{\it exponential} radial number density distributions and indeed are
recovering the observed dependence of the cluster dimension on the
number of lumps. In particular, accepting a linear regression we
obtain $D_{simul}=0.0013 N_{lumps}+1.471$. Actually, a function
converging asymtotically toward $D$=2 for large lump numbers would be
more appropriate; having no clue as to its exact form, though, we stay
within the linear approximation. Then, second, instead of using the
observed cluster dimensions as inferred from galaxy images, we
introduce {\it reduced cluster dimensions} defined by $D\equiv
D_{obs}-0.0013  N_{lumps}$, i.e. all measured cluster dimensions are
made comparable by formally adjusting them to the common number
$N_{lumps}$=0. Other values could have been chosen; however, the
adopted value (or other low values, say
$N_{lumps}$$\stackrel{_<}{_\sim}$50) yields consistently cluster
dimensions of about or below the theoretical maximum value of two. We
show in Fig.\/7, right, the reduced cluster dimension as a function of
the extrapolated central surface brightness for all galaxies. There is
a weak but significant {\it trend that fainter dwarf galaxies exhibit
lower cluster dimensions, i.e. more strongly clustered star-forming
regions, than brighter dwarf galaxies}. The same statement holds if
instead of central surface brightness we take the absolute magnitude
of the galaxy.

We have also determined the cluster dimensions for 15 selected
sub-galactic areas (consisting of about 30 lumps within a circle of
about 1.5 kpc diameter) within larger galaxies (each with a total of
more than about 200 lumps). With a mean of $D$$\approx$1.85 and a
scatter of only about 0.1, these areas show relatively high cluster
dimensions that are typically lying {\it above} their galaxies'
values. It furthermore implies that cluster dimensions for local 
lump aggregates scatter less than those for entire galaxies.\\

We now attempt to give an interpretation of the reduced cluster
dimension in terms of intragalactic gas porosity and star formation
rate. The volume filling factor $f$ of the {\it empty or low-density}
regions of a self-similar medium, the porosity, can be related to the
medium's fractal dimension in three dimensions, $D_3$, by
\[ f=1-\left(\frac{r_l}{r_u}\right)^{3-D_3}\,,\] where $r_l$ and $r_u$ 
are the lower and upper boundary of the relevant scaling range (e.g.,
Turcotte 1992). From Fig.\/7, left, and as mentioned above, we infer
$r_l\approx 100$ pc and $r_u\approx 1000$ pc. This approach to galaxy
porosity is analogous to Elmegreen's (1997) treatment of fractal
interstellar gas clouds, the porosity of which was characterized by
$f_{ICM}=1-C^{(D_3/3)-1}$, with a maximum density contrast of
$C\approx10^3-10^4$ for the intracloud gas. The two approaches are
formally and numerically similar if we identify $C=(r_u/r_l)^3\approx
10^3$. Qualitatively, dwarf irregular galaxies may thus be considered
as huge star-forming clouds similar to fractal intragalactic
star-forming clouds. Solving for the dimension, we obtain
\begin{eqnarray}D_3 &\approx& 3 + log(1-f)\,.\end{eqnarray}
Interpreting Fig.\/7, right, in terms of galaxy porosity, we have to
take into account that the scaling dimension of a {\it projected}
isotropic self-similiar object is one less than the true dimension
(Elmegreen \& Elmegreen 2001), thus $D_3=D+1$. We then learn that {\it
on average} fainter galaxies with {\it on average} lower cluster
dimensions, i.e. with stronger clustering properties, are also more
porous ($D\approx1.5$, $f\approx0.7$) than brighter galaxies
($D\approx1.9$, $f\approx0.2$).\\   Theoretically porosity is thought
to be crucial for the self-regulation of disks, and one expects an
increasing star-formation rate to be accompagnied with decreasing
porosity (Silk 1997, equ. 7). This holds empirically as well, as we
will sketch now. For dwarf irregular galaxies the area-normalized star
formation rate is correlated with the galaxy's extrapolated central
surface brightness: from Fig. 7a in van Zee (2001) we infer
$\mu_B^0\approx -1.79_{\pm 0.18}\,\log(SFR/area) + 18.214_{\pm
0.334}$, with $area\equiv\pi (1.5 R_d)^2$ and $R_d$ being the
exponential-model scale length in kpc. On the other hand, an ordinary
least-squares bisector fit (Isobe et al. 1990) to the data of
Fig.$\,$7, right, yields $\mu_B^0= -3.25_{\pm 1.01}\,D + 26.862_{\pm
1.560}$, shown as line in the figure. Equating the two expressions,
inserting equation (3), and remembering $D=D_3-1$, we finally deduce
\begin{eqnarray}
SFR\,[\mathrm{M}_\odot \mathrm{yr}^{-1}] &\approx& 
0.45\,(1-f)^{1.8}\,(R_d\,[\mathrm{kpc}])^2\,.
\end{eqnarray} 
Within our model treatment of dwarf irregular galaxies being
self-similar objects we thus have semi-empirically established a {\it
statistical} relation between SFR, scale length, and porosity, in the
sense that for a given scale length galaxies with higher SFRs are also
less porous. Note that for a given scale length, equation (4) predicts
a  maximum SFR. However, porosity as defined above has to be
understood as  a conceptual parameter and not as a quantity describing
reality in detail. The parameters possibly influencing the mean
porosity of a galaxy are manyfold (gas density, gas pressure or
velocity dispersion, gas metallicity,  supernova energy release),
forming an intricate, interdependent parameter set  (Silk 1997).

\section{Comparing clustering, spreading, and relative luminosity of the lumps}

\begin{figure}[t] 
\begin{center} 
 \includegraphics[width=85mm]{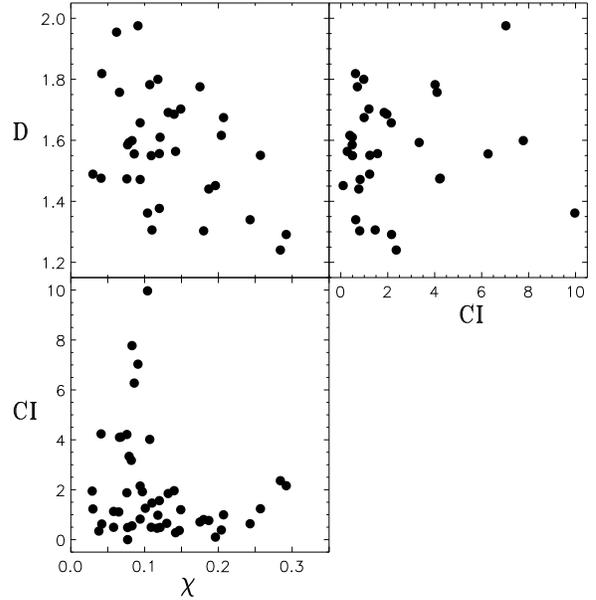} 
\end{center} 
\caption[]{Comparison of three morphological indices, as applied to
bright lumps within irregular dwarf galaxies: the lumpiness index
\Chi, the normalized concentration index $CI(x)$ (where
$x=R_{25}/R_d$), and the reduced cluster dimension $D$ (taking into
account the effect of differing disk scale lengths).}
\label{chi_CI_D}
\end{figure} 

In the course of this paper we encountered three indices, namely the
lumpiness index \Chi (Section 3), the concentration index $CI$
(Section 4.3), and the cluster dimension $D$ (Section 5). The first
one has been introduced only, but has not been applied until now. To
make up leeway and to draw some first conclusions, we plot in
Fig.$\,$\ref{chi_CI_D} the values resulting from applying the three
indices against each other. Several distinctive features are
present. First, and most notable, the lumpiness index \Chi$\,$
anticorrelates with cluster dimension (upper left panel): {\it the
higher the lumpiness index , i.e. the higher the fraction of total
galaxy light in the lumps, the smaller the cluster dimension, i.e. the
more clustered are the bright lumps}. Without plotting the
corresponding graphs, we note that the lumpiness index shows no
dependence on galaxy luminosity; this is contrary to the behaviour of
the cluster dimension. In terms of porosity, this behaviour means that
{\it lumpier galaxies are also more porous, and vice versa}.

Second, the degree of lumpiness and concentration behave such that the
galaxies with the highest values for the lumpiness index all have low
values for the concentration index (lower panel). In other words, high
fractional lump luminosities come with star-forming complexes that are
widely scattered within a galaxy disk.  On the other hand, those
galaxies with lumps very centrally concentrated, i.e. with $CI$$>$3,
come with low fractional lump luminosities, i.e. with low \Chi
values.  This is somewhat surprising because there are BCD- and
BCD-like galaxies with {\it central} starbursts (high $CI$) that are
expected to be very lumpy (high \Chi) at the same time. But indeed,
{\it  no galaxy with high fractional lump luminosity is observed, not
even a BCD galaxy, to be centrally  concentrated. This means that even
for actively star-forming galaxies the main body of light is still
clearly dominated by the total galaxy light.} We disclaim a selection
bias to  the disadvantage of dwarfs with central starbursts in the
next section by providing  two representative examples.

Finally, no relation seems to hold between concentration index and
cluster dimension (right panel). In particular, galaxies with highest
lump concentrations present any cluster dimensions. In other words,
lump location and the degree of self-similar clustering are
independent of each other; this argues, as done in  Section 4, for the
introduction of some {\it mean} galaxy porosity that may vary among
galaxies.

\section{Discussion and conclusions}

Regarding the azimuthally integrated, radial distribution of bright
lumps --- corresponding to star-forming complexes --- in dwarf
irregular galaxies, we find them non-uniformly distributed. While in
individual galaxies the number distribution is non-monotonic and
rugged, the summed-up distribution for all galaxies of our sample
manifests the hidden constraint, which is a $r\,e^{-r}$-distribution
closely tracing the underlying older population.  More precisely, in
terms of radial number density distribution the lumps follow an
exponential decay with scale length about 10 percent smaller on
average than that of the blue continuum light. This is consistent with
studies of the radial distribution of H$\,$II regions in
intermediate-type spiral galaxies (Athanassoula et al. 1998).
{\it The fact that each component of the average disk --- from
star-forming site number density to surface brightness of the total
light --- is approximately exponential is a hint that luminous
exponential disks are born rather than made, consistent with the
accretion scenario for the viscous evolution of galaxy disks} (e.g.,
Ferguson \& Clarke 2001).  Star-forming complexes in irregular dwarf
galaxies can be found out to large radii, as already emphasized by
Schulte-Ladbeck \& Hopp (1998), Brosch et al. (1998), and Roye \&
Hunter (2000). {\it The presence of a tail in the accumulated
radial number distribution of star forming regions out to at least six
optical scale lengths (Fig. 4)  indicates that the distributions of
dwarf irregulars are truncated at rather low gas density thresholds
for star formation. (van Zee et al. 1997, Hunter et al. 1998, Pisano
et al. 2000); this seems to be different with many spiral galaxies for
which sharp to weak truncations, starting at galactocentric radii of
2-4 near-infrared or 3-5 optical scale lengths, have persistently been
reported (e.g., recently, Florido et al. 2001, Kregel et al. 2002,
Pohlen et al. 2002).}

Beside the presence of main or primary peaks in the radial lump number
distribution at slightly less than one optical scale length on
average, there is --- contrary to the expectations for
exponential-disk systems --- the  frequent occurence of secondary
peaks at about two scale lengths. As simple simulations show, it is
consistent with the idea of triggered star formation based on a
stochastic self-regulation scenario. However, some of the brighter,
larger galaxies exhibit pronounced primary peaks at two scale lengths
and show  minor, secondary peaks around one scale length. For these
galaxies with a reversed peak pattern the simulations indicate that
shear-induced star formation around the disk's turnover to
differential rotation could be at work; we feel this issue worth a
deeper investigation: given a lump statistics based on
higher-resolution images and linked to detailed rotational velocity
data, and possibly supplemented with information on large-scale
magnetic field structures, this may lead to some subtle but decisive
insights related to star formation in irregular galaxies. Also,
the possible role of bars or bar-like central features should be
carefully considered.  However, as none of our four galaxies with
principal peaks appering around two scale lengths is classified as
``barred'', and because Roye \& Hunter (2000) did not see a
preferential location of H$\;$II regions towards the ends of bars in
the two candidate galaxies of their sample, we do not consider this
mechanism as being effective in shaping the number distribution of
lumps.

The observation that the scale lengths are the larger the older the
underlying respective population is, goes in hand with the finding
that in star-forming dwarf galaxies the oldest populations are also
the most extended ones (Gardiner \& Hatzidimitriou 1992, Minniti \&
Zijlstra 1996, Minniti et al. 1999, Harris \& Zaritsky 1999). The
common interpretation is that of an age-related dispersion of
stars. Because dynamical disk heating tends to saturate at a fixed
velocity dispersion (Freeman \& Bland-Hawthorn 2002) the amount of
radial spread introduced by dynamical heating  is expected to be
larger in smaller galaxies with lower gravitational binding energies
(J. Gallagher, private communication). Indeed, the fact noted in
Parodi et al. (2002) that with increasing scale length the disk color
gradients become systematically less positive and even  start being
weakly negative when going from dwarf irregular to low-surface
brightness and spiral galaxies ---  which is equivalent to larger
galaxies having red-to-blue band scale length ratios below one ---
seems to support the idea that the extent of the red component
decreases as galaxy mass increases. However, for spirals part of this
colour trend is most certainly a metallicity effect.

We provide no investigation of the azimuthal lump distribution; where
necessary we assumed axisymmetry.  Applying different asymmetry
indices to dwarf and normal irregulars, Heller et al. (2000) and Roye
\& Hunter (2000) obtained opposing results that were dependent on the
definition of the chosen index (asymmetry as a question of
perception). Also, it was found that while the conveniently applied
rotational asymmetry index may be used as a first discriminator
between (distant) elliptical and spiral/irregular galaxies (Schade et
al. 1995, Conselice et al. 2000), this parameter actually is strongly
dependent on recent star formation (Takamiya 1999, Mayya \& Romano
2001) and thus must be  correlated with the lumpiness index as adopted
in the present paper. As we have no data  for the star formation
rate of our galaxies, an immediate comparision cannot be offered,
however.

Applying a concentration index $CI$ that is normalized according to
the exponential-disk structure of a mean lump distribution leads to
consistent results for varying aperture sizes. It may also remove the
discrepancy found by Heller et al. (2000) between the $CI_{H\alpha}$
values for actual galaxies and the lower ones for simulated galaxies
with random  star formation region positions. Roye \& Hunter (2000)
pointed out an increased scatter of concentration indices for faster
rotating galaxies of their sample; we no longer see this effect with
our larger sample.

Comparing concentration ($CI$) with lumpiness (\Chi) we find the
galaxies with a high percentage ($>10\%$) of light stemming from the
lumps showing low to moderate concentrations
($CI\stackrel{_<}{_\sim}2$), i.e., galaxies with lumps that are widely
scatterd within the disk  maintain a higher fraction of the total $B$
luminosity (Fig. \ref{chi_CI_D}). On the other hand, for very
concentrated galaxies ($CI\stackrel{_>}{_\sim}2$) less than about 10
percent of the light is due to the lumps; actually, for these galaxies
the values for the lumpiness index $\Chi$ {\it fluctuate} around the
sharp value of 7\% attributed to most of the (barred) spiral and
irregular galaxies observed by Elmegreen \& Salzer (1999). They
suggested the similarity of the blue-band light fraction in complexes
for several galaxies of different Hubble types and different total
luminosities being due to similar star formation efficiencies.

It remains, however, remarkable that galaxies with very high lump
concentrations are {\rm not} among the galaxies showing high $B$
luminosity fractions of the lumps (Fig. 8, bottom panel).  One may
wonder whether some of the nearby, well-known BCDs with central
starbursts would agree with this conclusion as well, i.e. whether their
central bursts should not dominate the total light content. We
therefore examined where the typical starburst dwarf irregular
galaxies NGC$\,$1569 and NGC$\,$1705 would fill in the $CI$ vs. \Chi
diagram. Based on the 48 brightest central star clusters of
NGC$\,$1569 as compiled in Hunter et al.  (2000) and adopting a
distance of 2.5 Mpc and a galaxy absolute magnitude of $M_V=-17.99$
mag, we estimate a mere $\Chi \le 0.10$ for NGC$\,$1569. Similarly,
for the super-star cluster dominated amorphous galaxy NGC 1705, the
absolute magnitude for the 7 brightest clusters is  about $M_V=-14.1$
according to the data in O'Connell et al. (1994), whereas the galaxy
has a magnitude of $M_V=-16.13$ at a distance of 5 Mpc, implying only
$\Chi \approx 0.15$. Thus independent of the exact concentration
indices for the brightest clusters, these two small but representative
galaxies with central starbursts would not occupy the empty part of
the $CI$ vs. $\Chi$ diagram. This means that the empty region in this
figure is not a selection effect, but may be related to our adopted
procedures in determining the corresponding indices.

While the concentration index is a measure for lump spreading, the
cluster or correlation dimension provides information on the scaling
behaviour for lump-to-lump distances. We found the lump cluster
dimensions --- corrected for the effect of radial abundances according
to the annulus-integrated exponential distribution --- to lie between
1.3 and 2.0 and to gently correlate with extrapolated central surface
brightness and absolute magnitude of the host galaxy. At the same
time the cluster dimension is weakly anticorrelated with the
lumpiness index.

\begin{figure*}[t]
\hspace{-5mm}
 \includegraphics[width=185mm]{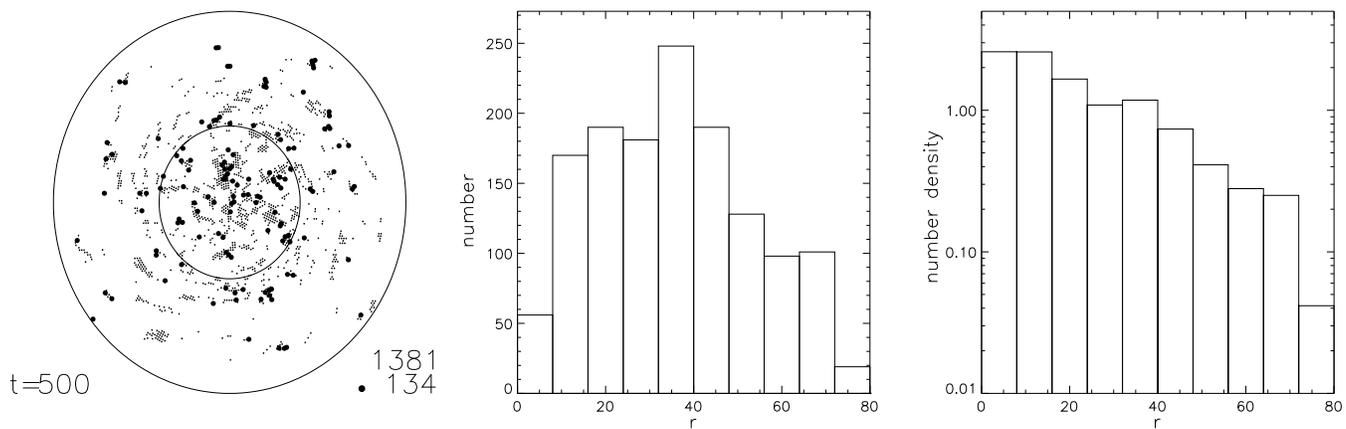} 
\caption{SSPSF simulation of an exponentiated disk with 80 corotating
rings and a transition from rigid to strong differential rotation at
two scale lengths (ring 32, solid line), after 500 time steps. The
three panels show the pattern of newly activated cells (filled
circles) together with those created less than 10 time steps ago
(dots), the radial number distribution of the active cells (bin width
is 8 rings or half a scale length), and the corresponding number
density distribution. Note that for the number distribution the peak
at one scale length (ring 16) --- the presence of which is expected
for non-differentially rotating exponential disks --- is markedly
surpassed by a peak around two scale lengths.}
\label{sspsf_galaxy}
\end{figure*} 

Cluster dimension (or porosity) as introduced in this paper may be
intimately linked to the sizes of the largest, kiloparsec-sized lump
compounds (Elmegreen et al. 1996) or to the sizes of star-forming,
collapsed expanding shells (e.g., Walter 1999). Elmegreen et
al. (1996) found the sizes of the largest compounds within spiral and
irregular galaxies to approximately scale with the square root of the
galaxy luminosity, or, if normalized by the galaxy semi-major axis
$R_{25}$, small galaxies to have slightly larger relative compound
diameters than larger galaxies. This was hypothesized to result from
gravitational instabilities with the Jeans length or the mean virial
density varying with galaxy luminosity. Walter (1999) suggests the
holes in dwarf irregulars to be larger than those in late-type spirals
because small galaxies have lower masses and correspondingly lower
gravitational potentials and lower ambient ISM gas densities, favoring
H$\,$I shells to grow larger. Alternatively, we are temptatively
interpreting the cluster dimension of a galaxy in terms of the volume
filling factor of empty regions in a fractal medium, i.e. in terms of
porosity (defined analogous to Elmegreen 1997), and find the following
statistical trends: {\it (i) fainter galaxies tend to be more porous;
(ii) more porous galaxies also have a lumpier  morphology with lower
central lump concentrations; (iii) the more porous the galaxy, the
lower the star formation rate per kpc$^2$ (equation 4). While these
trends are not unexpected, we  provide an objective and  quantitative
statistical treatment of these}.

Porosity, or self-similarity, as observed with the bright-lump
distribution within dwarf irregular galaxies reflects the
self-regulated evolution of the interstellar medium, with stellar
feedback and self-gravitation being the main mechanisms. It is thus
not to be confused with the still self-similar pattern of dispersed
stellar aggregates that initially formed from the fractal interstellar
gas, obeying the canonical value of $D$=1.3 for turbulence-driven star
formation (Elmegreen \& Elmegreen 2001). Our sub-galactic areas are
showing $D$$\stackrel{_>}{_\sim}$1.7, i.e. much higher fractal
dimensions. We suspect that in dwarf irregular galaxies either the
dispersive redistribution of stars is indeed much more effective than
in spiral galaxies (Elmegreen \& Hunter 2000), and/or that feed-back
regulation is responsible for the partially randomized position
patterns (e.g., due to the intersection of giant shells).\\ 

\begin{acknowledgements}
B.R.P. thanks Helmut Jerjen for introducing him to IRAF and for
pointing out the HIPASS data release interface. Valuable
recommendations by the referee, J. Gallagher, helped to improve the
paper. Financial support by the  Swiss National Science Foundation is
gratefully acknowledged.
\end{acknowledgements}

\appendix
\section{The SSPSF model}

The implementation of the stochastic self-propagating star formation
(SSPSF) disk model is a variant of the prescriptions found in the
reviews of Schulman \& Seiden (1986) and, in more detail,
Seiden \& Schulman (1990). Major modifications are an exponentiated
initial disk structure, a transition from a linearly increasing to a
flat rotation curve, and the inclusion of spontaneous star
formation. In particular, our model consists of 80 corotating rings,
each with 6\/R cells, R being the ring number or radius. Thus we have
a total of 18960 cells. Initially, for each ring the number of
occupied cells, $N(r)$, has a probability proportional to a Gamma
distribution $R\,exp(-R/R_d)$, with the scale length taken to be
$R_d=16$. On average the galaxies of our sample have a scale length of
about 0.7 kpc, thus the linear size of each cell corresponds to about
44 pc. The polar angles for the $N(r)$ cells in a ring are randomly
chosen. At the start there are a total of 300 occupied cells. The life
time of an occupied cell is 10 time steps, corresponding to about 10
Myr. The first time step following the activation of a cell, an empty
neighbouring cell has a probability of 0.21 to become activated
too. This implements the idea that the stellar wind of massive stars
created in a cluster travels with a wind velocity of about 40
km$\,$s$^{-1}$ for about 1 Myr forming an increasingly dense shell
that eventually fragments or hits other overdense regions, thus giving
chance to the formation of new star-forming sites. At each time step,
an additional 20 new cells are spontaneously activated; this sustains
the number-density profile being exponential. Typically, an
equilibrium occupation of around 1400 cells is reached after a few
dozen time steps (providing a filling factor around 0.07), 10 percent
of which having been just activated, and with the number density
distribution remaining rather exponential out to about four scale
lengths; further out there is a rapid drop or truncation of
star-forming regions.  \\ Either the rings rotate rigidly, with
circular velocity being proportional to the ring radius, or
differential rotation may be imposed by additionally demanding a
constant circular velocity beyond a turnover radius $R_t=2 R_d$. For
the simulations that included shear a flat rotation curve velocity of
500 km$\,$s$^{-1}$ was enforced, which is an unrealistic factor of
$\sim$10 faster than observed for a typical dwarf irregular. However,
this was only to clearly demonstrate the effect of strong shear within
the {\it model}. As long as the circular velocity is much higher than
the radial propagation velocity, shear seems to increase the star
formation rate. For each simulated galaxy the number (and number
density) distribution after 500 time steps was stored for use in the
study of Section 4.4. The lump patterns and their corresponding
distributions for a typical simulated galaxy can be seen in Fig. 9.
Note that we simulate only the occurence of new and the presence of
young stellar clusters, but  otherwise assume a prevailing underlying
population of older stars.

\end{document}